# A new approach to solar flare prediction

**Michael L. Goodman**[1,†], **Chiman Kwan**[2], **Bulent Ayhan**[2], **Eric L. Shang**[2]

[1] *Jacobs Space Exploration Group, Natural Environments Branch-EV44, NASA Marshall Space Flight Center, Huntsville, AL 35812, USA*
[2] *Applied Research LLC, 9605 Medical Center Drive-Suite 127 E, Rockville, MD 20850, USA*
*Corresponding author. E-mail:* †*michael.l.goodman@nasa.gov*

All three components of the current density are required to compute the heating rate due to free magnetic energy dissipation. Here we present a first test of a new model developed to determine if the times of increases in the resistive heating rate in active region (AR) photospheres are correlated with the subsequent occurrence of M and X flares in the corona. A data driven, 3D, non-force-free magnetohydrodynamic model restricted to the near-photospheric region is used to compute time series of the complete current density and the resistive heating rate per unit volume [$Q(t)$] in each pixel in neutral line regions (NLRs) of 14 ARs. The model is driven by time series of the magnetic field $\boldsymbol{B}$ measured by the Helioseismic & Magnetic Imager on the Solar Dynamics Observatory (SDO) satellite. Spurious Doppler periods due to SDO orbital motion are filtered out of the time series for $\boldsymbol{B}$ in every AR pixel. For each AR, the cumulative distribution function (CDF) of the values of the NLR area integral $Q_i(t)$ of $Q(t)$ is found to be a scale invariant power law distribution essentially identical to the observed CDF for the total energy released in coronal flares. This suggests that coronal flares and the photospheric $Q_i$ are correlated, and powered by the same process. The model predicts spikes in $Q_i$ with values orders of magnitude above background values. These spikes are driven by spikes in the non-force free component of the current density. The times of these spikes are plausibly correlated with times of subsequent M or X flares a few hours to a few days later. The spikes occur on granulation scales, and may be signatures of heating in horizontal current sheets. It is also found that the times of relatively large values of the rate of change of the NLR unsigned magnetic flux are also plausibly correlated with the times of subsequent M and X flares, and spikes in $Q_i$.

**Keywords** active regions, magnetic fields, flares, forecasting, heating, photosphere, models, magnetohydrodynamics

Contents

## 1 Introduction

Flares are concentrated in neutral line regions (NLRs) of active regions (ARs). NLRs are sites of relatively large cur-

*arXiv: 2003.01823.

rent densities that represent free magnetic energy available for conversion into particle energy during the flaring process [1–6]. Coronal observations show that for M and X flares, smaller pre-cursor flares occur in the same region within ∼ 24 hours prior to the main flares [7], and that there are temporal and spatial correlations between consecutive flares [8]. These facts suggest that, in the corona, the NLR current system evolves towards a large flaring event, and that this evolution involves smaller scale energy releases via current dissipation. Observations of quiet Sun network magnetic fields in the transition region indicate the appearance and intensification of non-potential magnetic fields, and hence of current densities, and suggest this can lead to network scale, magnetic reconnection driven eruptive events similar to, though on a smaller scale than coronal flares [9]. This process is observed to be correlated with photospheric magnetic flux emergence, and might be causally connected with coronal flaring [9]. Simulations and observations indicate that forming ARs expand into the corona with net currents that concentrate in NLRs [5, 10]. Collectively, these coronal and transition region observations and simulations raise the questions of whether there are corresponding current enhancements and heating events in the photosphere, and whether they are correlated with coronal flaring.

Observations of the complete photospheric current density $\boldsymbol{J}$ at sufficiently high spatial and temporal resolution and coverage can answer these questions. Computation of $\boldsymbol{J} = c\nabla \times \boldsymbol{B}/(4\pi)$, where $\boldsymbol{B}$ is the magnetic field and $c$ is the speed of light, requires knowing the 3D field $\boldsymbol{B}(x, y, z, t)$ at the photosphere, modeled here as a plane surface using Cartesian coordinates $(x, y, z)$, with $z$ height above the photosphere, and $t$ the time. Accurate computation of $\boldsymbol{J}$ from observations of $\boldsymbol{B}$ is challenging because its components are differences of derivatives of components of $\boldsymbol{B}$. This double differencing operation on components of $\boldsymbol{B}$ amplifies measurement error in $\boldsymbol{B}$.

There are models based on incomplete observations of the photospheric $\boldsymbol{B}$, such as those based on Michelson Doppler Imager (MDI, [11]) line-of-sight (LOS) magnetograms, and Imaging Vector Magnetograph magnetograms (IVM, e.g., [12]) that suggest affirmative answers to these questions, and show some effectiveness in predicting M and X flares and coronal mass ejections (CMEs).[1)] Examples of such models are those of Falconer *et al.* [13–17], Schrijver [18], Korsós *et al.* [19–22], Georgoulis & Rust [23] (also see [24]), and Leka & Barnes [25–27] and Barnes & Leka [28] (also see [1, 29, 30]). A review of some of these models, and of others, along with a comparison of their effectiveness in predicting flares is given by Barnes *et al.* [31], in which it is concluded that the models considered do not perform substantially better than climatological forecasts, which are based on long-term averages.

It is claimed here that the predictions of the better performing models are more directly based on estimates of components of $\boldsymbol{J}$, and that there is a positive correlation between larger values of $J$ and the occurrence of M and X flares. Of the 11 flare prediction algorithms reviewed and compared in Barnes *et al.* [31], none are found to be clearly superior to others, and none are found to be especially effective. However, the algorithms of Schrijver [18] and Falconer *et al.* [13–17] are among the top performing algorithms. It is claimed here that the reason for this is that those algorithms come closest to directly computing part of the current density, and using its magnitude as a major component of the algorithm. Due to the fact that observations do not provide the complete photospheric $\boldsymbol{B}$, none of these models compute the complete photospheric $\boldsymbol{J}$. The success of these models suggests that greater success in forecasting flares may be achieved by using models that: (i) Compute the complete $\boldsymbol{J}$. (ii) Are driven by higher resolution measurements of $\boldsymbol{B}$ that cover entire ARs continuously in time. (iii) Do not constrain the model-derived $\boldsymbol{B}$ to be force-free, so that photospheric phenomena, such as NLR current sheet dynamics, which involves non-zero Lorentz forces and heating by currents $\perp\boldsymbol{B}$, can be extracted from the data through the models.

### 1.1 An HMI data driven model that computes the complete photospheric current density

A model of this type is presented here, along with initial results based on the analysis of time series of $\boldsymbol{B}$ from 14 ARs. The model is driven by the 2D photospheric $\boldsymbol{B}(x, y, t)$ observed by the Helioseismic & Magnetic Imager (HMI, [32–34]) on the Solar Dynamics Observatory (SDO) satellite. The model combines this data with the $\nabla \cdot \boldsymbol{B} = 0$ condition to determine an analytic expression for a 3D magnetic field $\boldsymbol{B}(x, y, z, t)$ that is valid sufficiently close to the photosphere at $z = 0$. This allows the model to predict the complete $\boldsymbol{J}$ at the photosphere. HMI has the unique capability of combining full disk, continuous time observations of $\boldsymbol{B}$ with a spatial and temporal resolution of $1''$ and 12 minutes, which are sufficiently fine to begin to resolve the dynamics of granules. Granules are the smallest convection cells in the photosphere. They have diameters $d \sim 1000$–$2000$ km, and flow speeds $v \sim 2$–$7$ km·s$^{-1}$ [35]. Their lifetime is defined as the convective turnover time $\tau \sim d/v \sim 2.4$–$16.7$ minutes. Observations show that the photospheric magnetic field is concentrated in inter-granular lanes, and magnetic flux continually emerges into the photosphere in these regions, with quiet Sun field strengths up to several hG in internetwork, $\sim 10^3$ G in network, and with the emergence first appearing as a region of mixed polarity, nearly horizontal magnetic flux [36–40]. The photospheric $\boldsymbol{J}(x, y, 0, t)$ is obtained by analytically computing $\nabla \times \boldsymbol{B}(x, y, z, t)$, and

[1)] MDI provides full disk measurements of the LOS photospheric magnetic field with a spatial resolution of $4''$, and a full disk magnetogram temporal resolution of 90 minutes. IVM provides spatial and temporal resolutions $\sim 0.55'' - 2''$ and 2–4 minutes.

setting $z = 0$.[2)] This gives an estimate of the complete, full disk, photospheric $\boldsymbol{J}$ on granulation scales. The vector potential $\boldsymbol{A}$, and electric field $\boldsymbol{E}$ are also computed along with the resistive heating rate per unit volume ($Q$) at the photosphere. These quantities are used to compute $\boldsymbol{J} \cdot \boldsymbol{E}$ and $\boldsymbol{v} \cdot (\boldsymbol{J} \times \boldsymbol{B})/c$, where $\boldsymbol{v}$ is the center of mass (CM) velocity, which are respectively the rate per unit volume at which the electromagnetic (EM) energy and particle kinetic energy (KE) reservoirs exchange energy, and the rate at which the EM energy and CM KE reservoirs exchange energy.

The model and the computational algorithm that solves it using HMI data were developed on a NASA Phase 1 SBIR (Small Business Innovative Research) project (2014 - henceforth and in the bibliography referenced as S14)[3)]. The algorithm computes quantities in every AR pixel. Before computing quantities that involve $\boldsymbol{B}$, the spurious Doppler periods in the HMI measured $\boldsymbol{B}$ due to SDO orbital motion, and concentrated at periods of 6, 12, and 24 hours, are filtered out of the time series of $\boldsymbol{B}$ for each pixel using a Fast Fourier Transform (FFT) based band pass filter. Due to the time and funding constraints of S14, it was only possible to develop and implement the computational algorithm, and test it on 14 ARs using 3–4 day long time series of the HMI measured $\boldsymbol{B}$ for each AR.

### 1.2 Summary of the main results

Define $Q_i(t)$ to be the NLR area integral of $Q(x, y, 0, t)$. The algorithm used to select the pixels that define the NLR at each time for each AR is described in Section 2.3. The main results presented in this paper are:

i) Spikes in $Q_i$, appearing as increases by orders of magnitude above background values are found to occur in the NLRs of the 14 ARs. The largest spikes occur in the time series of $Q_i$ for ARs that have M or X flares, and their times of occurrence suggest it is plausible they are correlated with the occurrence of M or X flares a few hours to a few days later. A subset of these spikes was analyzed at the pixel level, and found to occur on the HMI and granulation scales of 1 arcsec and 12 minutes. Spikes are found in ARs with and without M or X flares, and outside as well as inside NLRs, but the largest spikes are localized in the NLRs of ARs with M or X flares, and associated with horizontal magnetic field strengths $\sim$ several hG, and vertical magnetic field strengths several orders of magnitude smaller. The large spikes in $Q_i$ are due to non-force-free currents $\boldsymbol{J}_\perp$ ($\perp \boldsymbol{B}$), and convection driven heating, which converts CM kinetic energy into thermal energy. At all times, and especially in strongly flaring ARs, resistive heating ($\propto J^2$) tends to be dominated by $J_\perp$ by orders of magnitude. These results suggest the large spikes are associated with horizontal, granulation scale current sheets. These results are discussed in Sections 5 and 6.

ii) Let $\Phi_i(t)$ be the NLR area integral of the signed magnetic flux through the photosphere. The times of occurrence of the larger values of $|\mathrm{d}\Phi_i/\mathrm{d}t|$ are plausibly correlated with the times of subsequent large spikes in $Q_i$, and M and X flares. A similar analysis of the time series of the unsigned magnetic flux, the magnitude of the signed magnetic flux, and the magnetic energy density does not suggest such correlations between their behavior and the occurrence of M or X flares, or spikes in $Q_i$. These results are discussed in Section 5.

iii) The model computed cumulative distribution function (CDF) of $Q_i$ for each of the 14 ARs is found to be essentially identical to the observed CDF of the energy $E$ of coronal flares. The CDF of $Q_i$ is the number $N(Q_i)$ of NLR area integrated heating rates with values $\geq Q_i$. Above an AR dependent value of $Q_i$, $N(Q_i)$ is found to be a scale invariant power law distribution, meaning $N(Q_i) = kQ_i^{-s}$ with $k$ an AR dependent number independent of $Q_i$, and $s$ independent of $Q_i$ and the same for all ARs to within about 1 standard deviation. The observed CDF $N(E)$ for coronal flares is $N(E) = KE^{-S}$ where $K$ is an AR dependent number independent of $E$, and $S$ is independent of $E$, and shows little statistical variation between ARs. It is found that the observed range of $S$ and the model computed range of $s$ are essentially the same. This similarity between $N(Q_i)$ and $N(E)$ suggests a connection between whatever process drives the photospheric $Q_i$, and the one that drives coronal flares. These results are discussed in Section 7.

## 2 The model

The model is a magnetohydrodynamic (MHD) model that uses all of the available HMI $\boldsymbol{B}$ data for a given AR. The model is valid close to the photosphere, meaning that the analytical, semi-empirically determined expression for $\boldsymbol{B}$ is valid through order $z^2$, and the $\nabla \cdot \boldsymbol{B} = 0$ condition is satisfied through order $z$, as discussed in Section 2.2. Once all derivatives with respect to $z$ are computed for quantities of interest, such as $\boldsymbol{J}$, the limit $z \to 0$ is taken to obtain the final set of photospheric time series. In this way, the 3D nature of the model differs from that of the many types of models that extrapolate the magnetic field from the photosphere into the corona (e.g., [41] and references therein).

The only differential equation in the model is for $\boldsymbol{A}$. This equation is solved analytically. $\boldsymbol{E}$ is determined from $\boldsymbol{A}$. There are no equations for density, tempera-

2) In the model, it is only at $z = 0$ that the $\nabla \cdot \boldsymbol{B} = 0$ condition is satisfied exactly. See Section 2.2.

3) A link to the final report for S14 is provided in the bibliography.

ture, pressure, or momentum. A data determined length scale $L(x, y, t)$ determines the derivatives of all components of $\boldsymbol{B}$ with respect to $z$ at the photosphere, allowing the complete $\boldsymbol{J}(x, y, 0, t)$ to be computed. It is enforcing the $\nabla \cdot \boldsymbol{B} = 0$ condition that allows $L(x, y, t)$, and hence $\boldsymbol{J}(x, y, 0, t)$ to be computed.

The HMI pixel side length is $0.5''$, with pixel overlap resulting in a magnetic field resolution $\sim 1''(\sim 725$ km). The HMI data pipeline provides time averaged data with a resolution of 12 minutes. Data with higher time resolution is available. The 12 minute averaged data from the hmi.sharp_720s_cea data series are used here to minimize effects of noise. Every 12 minutes HMI provides a full disk map of $\boldsymbol{B}$. HMI $\boldsymbol{B}$ is a function of $x, y$, and $t$. The use of the cea (cylindrical equal area) data helps minimize projection error, including correcting for the effect of solar differential rotation [34, 42].

The use of Cartesian coordinates and Cartesian magnetic field components in the model introduces projection error in the following sense. The hmi.sharp_720s_cea data series used here provides the magnetic field in spherical components $(B_r, B_\theta, B_\phi)$ as a function of spherical coordinates $(\theta, \phi)$ at the photosphere. $B_\phi$ is positive westward in the direction of solar rotation, $B_\theta$ is positive in the southward direction, and $B_r$ is positive upward from the photosphere. Here we set $B_x = B_\phi$, $B_y = B_\theta$, and $B_z = B_r$, with the $x, y$ directions chosen correspondingly. Therefore a transformation between spherical and Cartesian coordinates and vector components that accounts for the finite radius of curvature of the Sun is not used. The resulting error is projection error. The following estimates show this error is not expected to be large for ARs that are not too close to the limb where it is difficult to make accurate measurements. The maximum error is $\sim$ the ratio of the characteristic dimension of an AR ($\sim 7 \times 10^4$ km) to the solar radius $R_s$. This ratio is $\sim 0.1$. The error is expected to be smaller for our analysis since it is restricted to NLRs, which have areas smaller than those of their corresponding ARs. Pixel scale computations, such as of $\boldsymbol{J}$, have a much smaller error, on the order of $1''/R_s \sim 10^{-3}$ for HMI. In order to partially reduce projection error, the time series used here are chosen so the corresponding ARs stay within $\sim 60°$ of disc center during the duration of the time series.

## 2.1 The model magnetic field

The temporal and horizontal spatial variation of the photospheric magnetic field $\boldsymbol{B}$ is known down to the resolution of HMI. The HMI measurements are compromised by spurious Doppler oscillations, which we filter out of the data before using it to determine the model magnetic field. HMI does not provide information about the vertical dependence of $\boldsymbol{B}$. Since the horizontal $\boldsymbol{J}$ involves $\partial B_x/\partial z$ and $\partial B_y/\partial z$, it is necessary to know this dependence near the photosphere in order to determine the complete $\boldsymbol{J}$ at the photosphere. The model field is constructed to reproduce the HMI time series of $\boldsymbol{B}$ in each pixel in each AR, and to have a space and time dependent vertical variation at the photosphere determined by requiring that the $\nabla \cdot \boldsymbol{B} = 0$ constraint be satisfied exactly at the photosphere. The model field is not constrained to be force-free. These are natural requirements on the model field, but there are many ways to construct such a field. We choose a relatively simple form for this field, but do not claim it is an accurate representation of the actual photospheric field. Our objective is to find a way to improve flare prediction. The degree to which our choice of the model field is justified is the degree to which results derived from it improve flare prediction.

Let $L_x$ and $L_y$ be the $x$ and $y$ dimensions of the rectangular region used to enclose most or all of the AR modeled. The HMI pixel side length $\Delta = 0.5''$. The number of HMI data points covering this region is $N = (N_x + 1)(N_y + 1)$, where $N_x = L_x/\Delta$, $N_y = L_y/\Delta$, and $N_x$ and $N_y$ are given by the HMI datasets. For any function $f(x, y, z, t)$, define $f_{,x} = \partial f/\partial x$, and similarly for derivatives with respect to $y, z$, and $t$.

Let

$$\boldsymbol{B}(x, y, z, t) = \mathrm{e}^{-z/L(x,y,t)} \sum_{n=0}^{N_x} \sum_{m=0}^{N_y} \boldsymbol{b}_{nm}(t) \mathrm{e}^{2\pi \mathrm{i}\left(\frac{nx}{L_x} + \frac{my}{L_y}\right)}. \tag{1}$$

Here the $\boldsymbol{b}_{nm}(t)$ are complex, and $L(x, y, t) \equiv L_0(x, y, t) + zL_1(x, y, t)/L_0$ where $L_0$ and $L_1$ are real and determined by the HMI data and the $\nabla \cdot \boldsymbol{B} = 0$ condition, as described in Section 2.2. Equation (1) is assumed to be valid for sufficiently small $z$.

Equation (1) is chosen as the representation of $\boldsymbol{B}$ for the following reasons: (i) As discussed below, at $z = 0$ the 2D Fourier expansion of $\boldsymbol{B}(x, y, 0)$ allows the expansion coefficients to be determined rapidly using a FFT that uses all of the HMI data in each AR. The resulting $\boldsymbol{B}(x, y, 0)$ exactly reproduces the HMI measured field in each pixel. (ii) The data determines the space and time dependence of $L$, and so directly determines $\partial \boldsymbol{B}/\partial z$, and hence $J_x$ and $J_y$. (iii) The separable, exponential form of the $z$ dependence is chosen for simplicity. It allows $\partial \boldsymbol{B}/\partial z$ to be exactly computed analytically, as opposed to requiring numerical computation. Any determination of $\boldsymbol{B}$ is approximate, and the degree to which its mathematical representation is justified is determined by how useful it proves to be in understanding the solar atmosphere. Equation (1) appears to be a new representation of $\boldsymbol{B}$ at the photosphere. The results in this paper suggest it is useful.

For $z = 0$, and given the $N$ vectors $\boldsymbol{B}(x_i, y_i, 0, t_j)$ from the HMI data for each $j$, Eq. (1) represents $N$ complex, linear, inhomogeneous equations for the $N$ complex unknowns for each of the three components of $\boldsymbol{b}_{nm}(t)$. There are a total of $3N$ equations to solve at each time. The time series of the $\boldsymbol{b}_{nm}(t)$ are determined using an FFT algorithm to solve these equations. The imaginary part of $\boldsymbol{B}$ must be zero, which is used as a check on the numerical solution for the $\boldsymbol{b}_{nm}(t)$. Since $\boldsymbol{B}$ must be real, it is computationally efficient to work with the real part of Eq. (1). This is done as follows.

Let $\boldsymbol{z}_{nm} = \boldsymbol{b}_{nm}(t)\exp(2\pi \mathrm{i}(nx/L_x + my/L_y))$. Define its real and imaginary parts by

$$\boldsymbol{R}_{nm}(x,y,t) \equiv (\boldsymbol{z}_{nm} + \boldsymbol{z}^*_{nm})/2, \tag{2}$$

$$\boldsymbol{I}_{nm}(x,y,t) \equiv (\boldsymbol{z}_{nm} - \boldsymbol{z}^*_{nm})/(2\mathrm{i}). \tag{3}$$

Then $\boldsymbol{B}$ is re-defined as its real part, so now

$$\boldsymbol{B}(x,y,z,t) = \mathrm{e}^{-z/L(x,y,t)} \sum_{n=0}^{N_x}\sum_{m=0}^{N_y} \boldsymbol{R}_{nm}(x,y,t). \tag{4}$$

As discussed in Sections 2.3 and 2.4, $\boldsymbol{J}$ and $\boldsymbol{A}$ are determined analytically from $\boldsymbol{B}$, and $\boldsymbol{E}$ is determined by numerical time differentiation of $\boldsymbol{A}$. Since the magnetic field model reproduces the HMI determined field, errors in $\boldsymbol{J}$, $\boldsymbol{A}$, and $\boldsymbol{E}$ are due to inaccuracies in the data, which include limited time and space resolution, and due to any deviation of the actual height dependence of $\boldsymbol{B}$ at the photosphere from the assumed exponential factor $\exp(-z/L(x,y,t))$ where $z \to 0$, and $L$ is determined by the HMI data as discussed in Section 2.2. Because $\boldsymbol{J}$ involves differences of derivatives of components of $\boldsymbol{B}$, it is especially sensitive to measurement and model error. The magnetic field model is largely chosen on the basis of information available from observations of the field, which, relative to resolving the horizontal variation of $\boldsymbol{B}$, are severely lacking in resolution of the height dependence of $\boldsymbol{B}$. A major advance in the accuracy of computing the full $\boldsymbol{J}$ requires a major advance in resolving the height dependence of $\boldsymbol{B}$.

## 2.2 The $\nabla \cdot \boldsymbol{B} = 0$ condition

Define $\boldsymbol{B}_0 = \boldsymbol{B}(x,y,0,t)$. Take the divergence of Eq. (4) and set it equal to zero. Solving the resulting equation through order $z$ gives

$$L_0(x,y,t) = \frac{B_{0z}}{B_{0x,x} + B_{0y,y}}, \tag{5}$$

and

$$L_1(x,y,t) = -\frac{L_0}{2B_{0z}}(B_{0x}L_{0,x} + B_{0y}L_{0,y}). \tag{6}$$

The right hand sides of Eqs. (5) and (6) are evaluated at $z = 0$. Therefore, $L_0$ and $L_1$ are completely determined by the HMI data. The resulting expression for $\boldsymbol{B}$ is valid through order $z^2$. It is the $\nabla \cdot \boldsymbol{B} = 0$ condition plus the HMI data that determine the $z$ dependence of $\boldsymbol{B}$, which is what allows the complete expressions for $J_x$ and $J_y$ to be computed. Without a determination of the height dependence of $\boldsymbol{B}$, the components of $J_x$ and $J_y$ that involve $B_{y,z}$ and $B_{x,z}$ at $z = 0$ cannot be computed.

## 2.3 Current density

The current density $\boldsymbol{J}(x,y,z,t) = c\nabla \times \boldsymbol{B}/(4\pi)$. Through order $z$,

$$(\nabla \times \boldsymbol{B})_x = \exp(-z/L)\left[\frac{z}{L_0^2}L_{0,y}B_{0z} + B_{0z,y} + \frac{1}{L_0}\left(1 - \frac{2L_1 z}{L_0^2}\right)B_{0y}\right], \tag{7}$$

$$(\nabla \times \boldsymbol{B})_y = -\exp(-z/L)\left[\frac{z}{L_0^2}L_{0,x}B_{0z} + B_{0z,x} + \frac{1}{L_0}\left(1 - \frac{2L_1 z}{L_0^2}\right)B_{0x}\right], \tag{8}$$

$$(\nabla \times \boldsymbol{B})_z = \exp(-z/L)\left[\frac{z}{L_0^2}(L_{0,x}B_{0y} - L_{0,y}B_{0x}) + (B_{0y,x} - B_{0x,y})\right]. \tag{9}$$

For the ARs analyzed, the terms $B_{0x}/L_0$ and $B_{0y}/L_0$ in $J_y$ and $J_x$ give rise to spikes in the horizontal current density $J_h = (J_x^2 + J_y^2)^{1/2}$ at the photosphere that are orders of magnitude above background values, with the largest spikes occurring in the NLRs of strongly flaring (SF) ARs. Here an SF AR is defined as one with M or X flares, and a control AR (C AR) is defined as one with lower class or no flares.

The NLR for each AR at each time is determined in the following way. Schrijver [18] used MDI magnetograms to identify $3 \times 3$ pixel blocks that contain one or more pairs of opposite polarity LOS magnetic fields with magnitudes $> 150$ G. The set of such blocks contains the highest gradient polarity separation lines, which define the regions of highest current density. Each region is convolved with an appropriately centered Gaussian with a FWHM of 15 Mm centered on the region. The resulting set of pixels is defined to be the NLR. This NLR mapping algorithm was modified and implemented for HMI data by Bobra [43]. This modified algorithm is used here to define the NLR. The set of pixels that define the NLR varies with time, may comprise disjoint regions, and is re-computed by the model at each 12 minute time step. The NLRs analyzed here are comprised of $\sim 10^3$–$10^4$ pixels.

## 2.4 Vector potential and electric field

Assume the following expansion, valid through order $z^3$ for sufficiently small $z$.

$$\boldsymbol{A}(x,y,z,t) = \boldsymbol{a}_0(x,y,t) + \boldsymbol{a}_1(x,y,t)z + \boldsymbol{a}_2(x,y,t)z^2 + \boldsymbol{a}_3(x,y,t)z^3. \tag{10}$$

Expand $\boldsymbol{B}$ through order $z^2$ to obtain

$$\boldsymbol{B}(x,y,z,t)=\left(1-\frac{z}{L_0}+\left(1+\frac{2L_1}{L_0}\right)\frac{z^2}{2L_0^2}\right)\boldsymbol{B}_0(x,y,t). \tag{11}$$

The $\boldsymbol{a}_i(0\le i\le 3)$ are obtained by solving $\boldsymbol{A}=\nabla\times\boldsymbol{B}$ together with the Coulomb gauge condition $\nabla\cdot\boldsymbol{A}=0$, order by order in powers of $z$, for specified boundary conditions on the $\boldsymbol{a}_i$. The question is how to choose these boundary conditions. There is no guidance from observations for how to make this choice. Consequently, the choice is made heuristically as follows.

The electric field may always be written as

$$\boldsymbol{E}=-\frac{1}{c}\frac{\partial \boldsymbol{A}}{\partial t}-\nabla\Phi, \tag{12}$$

for some scalar function $\Phi$. The electric and magnetic fields are gauge independent [44]. The Coulomb gauge is the appropriate gauge to use for computing $\boldsymbol{E}$ in cases in which $\boldsymbol{E}$ is primarily an induction field, generated by $\partial\boldsymbol{B}/\partial t$, as opposed to an electrostatic field, generated by charge separation. The reason is that in the Coulomb gauge $\Phi$ is the electrostatic potential, so $-\nabla\Phi$ is the electrostatic field, in which case this term may be neglected in Eq. (12). This simplification is used here based on the assumption that the electrostatic field may be neglected in comparison with the induction electric field on the photospheric space and time scales of interest here, which are respectively $\gg$ the Debye length, and the inverse of the plasma and collision frequencies.[4)] This assumption is consistent with MHD because the quasi-neutrality approximation of MHD is by definition valid on these space and time scales. Then, in the Coulomb gauge, $\boldsymbol{E}\sim -c^{-1}\partial\boldsymbol{A}/\partial t$.

It follows that boundary conditions on $\boldsymbol{E}$ can to some extent be used to determine boundary conditions on $\boldsymbol{A}$. Assume that $E_z(x,y,0,t)=(\partial E_z(x,y,z,t)/\partial z)_{z=0}=0$ for all $t$. Consistent with this assumption, the choice $a_{0z}=a_{1z}=0$ for all $t$ is made. A heuristic justification for the assumed boundary conditions on $E_z$ is as follows. Treating the photosphere as a perfectly conducting surface implies $E_z=4\pi\sigma$, and $E_x=E_y=0$ at $z=0$, where $\sigma$ is the surface charge density. Assuming $\sigma=0$, and $\nabla\cdot\boldsymbol{E}=4\pi\rho_c=0$, where $\rho_c$ is the volume charge density, gives the assumed boundary conditions.

With these boundary conditions $\boldsymbol{A}$ is uniquely determined. Let $\Sigma'$ denote the sum over all $n$ and $m$ except the term with $n=m=0$. Then

$$A_x=-\frac{L_x^2L_y}{2\pi}\sum{}'\frac{mI_{z,nm}}{n^2L_y^2+m^2L_x^2}-\frac{yR_{z,00}}{2}+z\left(1-\frac{z}{2L_0}\right)B_{0y}+\frac{1}{6}\left[\left(1+\frac{2L_1}{L_0}\right)\frac{B_{0y}}{L_0^2}-(B_{0y,xx}-B_{0x,xy})\right]z^3, \tag{13}$$

$$A_y=\frac{L_xL_y^2}{2\pi}\sum{}'\frac{nI_{z,nm}}{n^2L_y^2+m^2L_x^2}+\frac{xR_{z,00}}{2}-z\left(1-\frac{z}{2L_0}\right)B_{0x}-\frac{1}{6}\left[\left(1+\frac{2L_1}{L_0}\right)\frac{B_{0x}}{L_0^2}-(B_{0x,yy}-B_{0y,xy})\right]z^3, \tag{14}$$

$$A_z=\left[-\frac{1}{2}(B_{0y,x}-B_{0x,y})\left(1-\frac{z}{3L_0}\right)+\frac{z}{6L_0^2}(B_{0x}L_{0,y}-B_{0y}L_{0,x})\right]z^2. \tag{15}$$

Direct calculation using this solution shows that $\nabla\cdot\boldsymbol{A}=0$ through order $z^2$.

### 2.5 Ohm's law

The assumed Ohm's law for the photosphere is

$$\boldsymbol{E}+\frac{\boldsymbol{v}\times\boldsymbol{B}}{c}=\eta\boldsymbol{J}. \tag{16}$$

Here $\eta$ is the parallel (Spitzer) resistivity, assumed constant with the value $2\times10^{-12}$ s (Fig. 3 of Ref. [45])[5)]. $\boldsymbol{E}$ and $\boldsymbol{J}$ can be computed once $\boldsymbol{B}$ is determined from the model.

Independent of the form of the Ohm's law, the rate $\boldsymbol{J}\cdot\boldsymbol{E}$ at which EM energy is converted into particle energy may be decomposed as $\boldsymbol{J}\cdot\boldsymbol{E}=Q+R_{CM}$. Here $Q\equiv\boldsymbol{J}\cdot(\boldsymbol{E}+(\boldsymbol{v}\times\boldsymbol{B})/c)$ is the Joule heating rate per unit volume, and $R_{CM}\equiv\boldsymbol{v}\cdot(\boldsymbol{J}\times\boldsymbol{B})/c$ is the rate at which EM energy is converted into CM KE. For the Ohm's law used here, $Q=\eta J^2$, and $R_{CM}$ is computed as $\boldsymbol{J}\cdot\boldsymbol{E}-Q$. Although the Ohm's gives $\boldsymbol{v}_\perp=c[(\boldsymbol{E}-\eta\boldsymbol{J})\times\boldsymbol{B}]/B^2$, $\boldsymbol{v}_\perp$ is not used in this paper.

If $|\boldsymbol{J}\cdot\boldsymbol{E}|\ll Q$ then $Q\sim -R_{CM}$. This is the case of convection driven heating, for which there is essentially no change in the EM energy of the plasma, and CM KE is transformed directly into thermal energy. In this case, the electric field that drives the heating is the convection electric field $\boldsymbol{E}_{conv}=(\boldsymbol{v}\times\boldsymbol{B})/c$. Noting that $\boldsymbol{J}\cdot\boldsymbol{E}=Q-\boldsymbol{J}\cdot\boldsymbol{E}_{conv}$, the case of convection driven heating may be defined by the condition $Q\sim\boldsymbol{J}\cdot\boldsymbol{E}_{conv}$. The convection electric field drives a current density $\boldsymbol{J}_\perp$ perpendicular to $\boldsymbol{B}$. As shown in Sections 5 and 6, the largest values of $Q_i$, which correspond to the strongest current enhancements, are almost entirely convection driven, while most smaller

4) The relevant Debye length and collision times are $\sim10^{-5}$ cm and $10^{-10}$–$10^{-9}$ s. These are insignificant compared with the HMI spatial and temporal resolutions of 1″ and 12 minutes.

5) This value of $\eta$ corresponds to a temperature $T\sim 6000$ K, and includes the effects of electron collisions with protons, HI, and HeI. See Ref. [45] for a detailed derivation of the conductivity tensor for a 3-particle species representation of the solar atmosphere that includes contributions from electrons, protons, HI, HeI, and singly charged ions other than protons.

values of $Q_i$ are not convection driven. Consistent with this result, it is shown that the larger values of $Q_i$ are due to increases in $J_\perp$ by orders of magnitude, while there is relatively little or no change in $J_{//}$, which is the magnitude of the current density parallel to $\boldsymbol{B}$.

Define $R_{CMi}$ as the NLR area integral of $R_{CM}$. It follows from $\boldsymbol{J} \cdot \boldsymbol{E} = Q + R_{CM}$ that $(\boldsymbol{J} \cdot \boldsymbol{E})_i = Q_i + R_{CMi}$. This relation is used to compute $R_{CMi}$ in Section 5 and 6 to show that the largest spikes in $Q_i$ are convection driven.

## 3 Removal of spurious Doppler periods from the HMI $B$ data

Due to the orbital motion of SDO, there are spurious, Doppler shift generated oscillations in the HMI data for the components of $\boldsymbol{B}$ in each pixel [33, 34, 46–49]. The fundamental oscillation period is 24 hours, but associated oscillations at 6 and 12 hours are observed, and other related spurious periods may also be present, although their values and ranges are not clearly known. The oscillation amplitudes can be as large as 100 G, with the largest amplitudes concentrated in strong field regions (i.e., $B > 1000$ G). Consequently, the resulting error in quantities that are sums over pixels of quantities derived from the magnetic field increases with the number of pixels, and can be very large. Since NLR integrated quantities typically involve sums over from $10^3$ to several times $10^4$ pixels, we decided to remove this Doppler noise by applying an FFT based band-pass filter to the time series of HMI B for each pixel. The filter removes all components of the magnetic field signal with periods $\geq$ 5.56 hours, but retains the DC component. This means our analysis is based on the high frequency component of $\boldsymbol{B}$, meaning that only periods $<$ 5.56 hours are retained in the signal, in addition to the DC component. The filtering probably eliminates all or almost all of the Doppler noise. But it also removes any real AR magnetic field dynamics with periods in this range. The HMI time series analyzed here have lengths of 70–140 hours, and AR magnetic fields typically undergo large changes on these time scales. The filter eliminates a large source of error due to the spurious Doppler periods, but may also remove multiple, real oscillations in $\boldsymbol{B}$ at periods $\geq$ 5.56 hours that occur during the duration of the time series. However, this elimination of part of the real magnetic field signal is not by itself a problem because the important question is whether the remaining high frequency component of $\boldsymbol{B}$, or quantities derived from it, prove useful for flare prediction. Based on results presented in Section 5, we claim it is plausible that the derived $Q_i$ and $|\mathrm{d}\Phi_i/\mathrm{d}t|$ are useful in this way, but a definitive statistical test of their usefulness must be based on an analysis of at least hundreds more HMI AR time series of $\boldsymbol{B}$. Henceforth it is understood that all magnetic field based quantities are high frequency quantities, meaning that they are derived from the high frequency $\boldsymbol{B}$ just defined. In Section 5.2 it is shown there is a plausible correlation between larger values of $|\mathrm{d}\Phi_i/\mathrm{d}t|$, spikes in $Q_i$, and M and X flares. Larger values of $|\mathrm{d}\Phi_i/\mathrm{d}t|$ correspond to larger values of $|\partial \boldsymbol{B}/\partial t|$, which correspond to the higher frequency component of $\boldsymbol{B}$. This suggests that filtering out the lower frequency component of $\boldsymbol{B}$, as is done here, retains the dynamics of $\boldsymbol{B}$ most important for flare prediction.

Examples of the effect of the filter are as follows. Denote the filtered and un-filtered time series for $B_x$ as $B_{xf}$ and $B_{xu}$, and similarly for other quantities. Figure 1 shows the filtered HMI time series of the components of $\boldsymbol{B}$, the difference between the un-filtered and filtered time series, and the magnitude of their ratio for a randomly selected pixel from the NLR of NOAA AR 1166 during a 70 hour long time series. This AR is one of the SF ARs analyzed here. The figure shows that the difference between the filtered and un-filtered time series is as large as 50 G. These differences are consistent with observations of the Doppler noise (see references at the beginning of this section). Similar differences are found for time series of $\boldsymbol{B}$ in other pixels that have maximum unfiltered field strengths up to a few hundred G. For pixels with larger unfiltered field strengths the differences tend to increase with field strength, reaching RMS values up to $\sim 10^2$ G. We believe these larger differences are due to the filtering out of some of the real AR dynamics with periods $\geq$ 5.56 hours, in addition to the Doppler noise.

As stated, the differences between filtered and unfiltered quantities that are integrals of pixel level quantities over NLRs can be large. For example, again consider the time series for NOAA AR 1166 used for Fig. 1. Figures 2–3 show the results of integrating the filtered and un-filtered pixel level results for $\eta J^2$ and $B^2/8\pi$ over the NLR at each time. The 70 hour long time interval includes 1 X, 2 M, and 9 C flares. For these and subsequent figures, the red, green, and light blue vertical lines and their labels indicate the times and magnitudes of X, M, and C flares. During the 70 hour long time series the number of pixels in the NLR varies across the range of $\sim 3 \times 10^4 - 6 \times 10^4$. The figures show large differences between the filtered and unfiltered quantities.

## 4 Fluctuations in $B$ and measurement error

There are three basic processes that can appear as random fluctuations in $\boldsymbol{B}$.

One is random fluctuations in the HMI measurement system. Their mean values are $\sim$ 100 G for $B$, with the noise in LOS $B \sim 10$ times smaller than in transverse $B$ [34, 50]. For the Cartesian representation used here, this approximately translates into mean noise levels $\sim$ 9.95 G in $B_z$, and $\sim$ 70.4 G in each of $B_x$ and $B_y$. Larger random fluctuations in $B$, up to $\sim$ 220 G, mainly in transverse $B$, are possible [33].

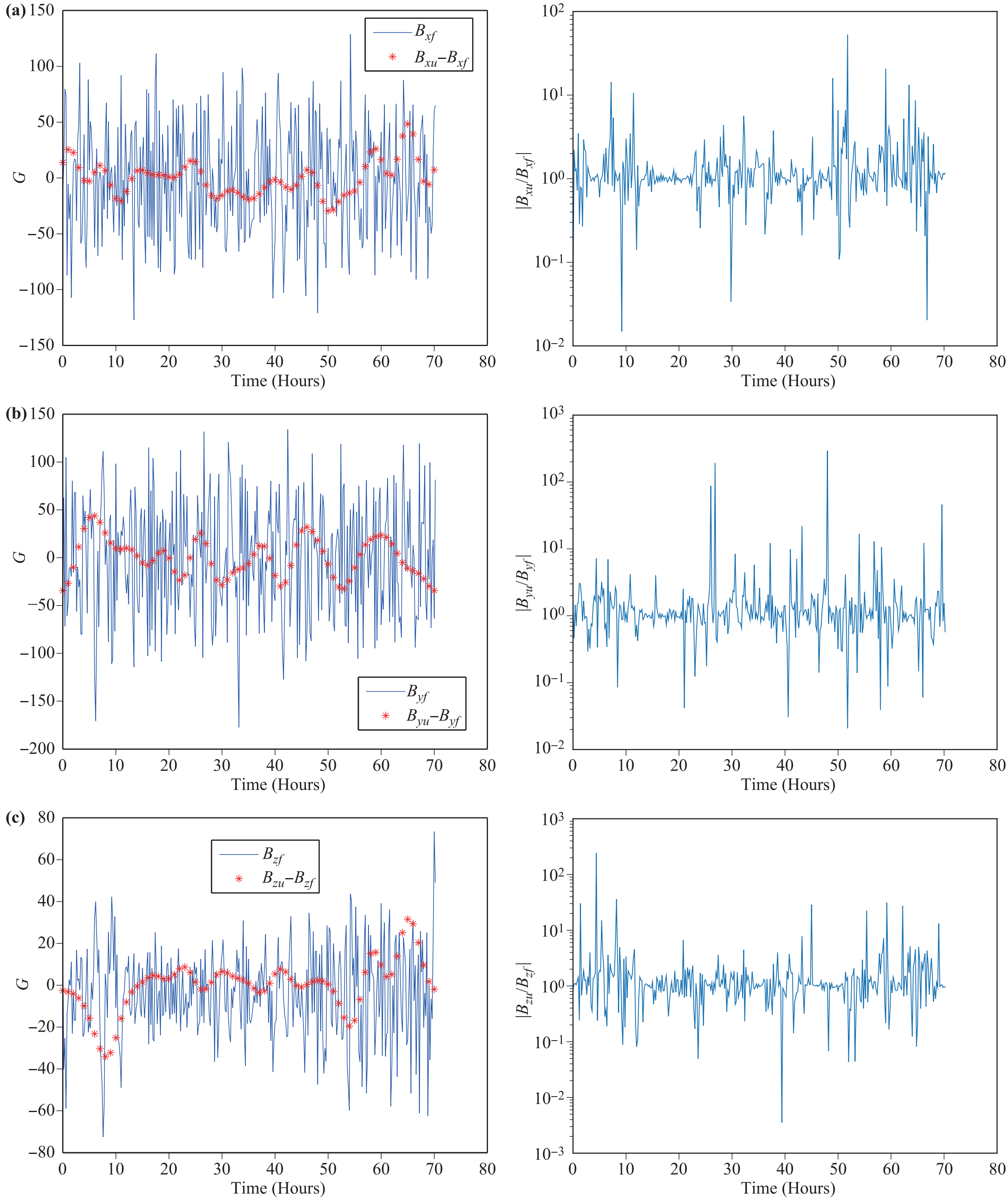

**Fig. 1** **(a–c)** Comparison of filtered and un-filtered $B_x, B_y, B_z$ in a pixel.

A second process is the random fluctuations in photospheric dynamics. This is part of the real dynamics of $\boldsymbol{B}$. The amplitude of these fluctuations is not accurately known. Here this amplitude is estimated using the observed approximate validity of the equi-partition condition $B = (4\pi\rho)^{1/2}v$, valid mainly for horizontal fields, but accurate to within a factor $\sim 2$ for stronger vertical fields [51]. Assume a constant background state with density $\rho_0$, velocity $\boldsymbol{v}_0$, and magnetic field $\boldsymbol{B}_0$, and random perturbations $\rho_1, \boldsymbol{v}_1$, and $\boldsymbol{B}_1$. The equi-partition assumption implies $(\boldsymbol{B}_0 + \boldsymbol{B}_1)^2/(8\pi) = (\rho_0 + \rho_1)(\boldsymbol{v}_0 + \boldsymbol{v}_1)^2/2$. Taking the time average $\langle\ \ \rangle$ of this expression gives $B_1^2/(4\pi) =$

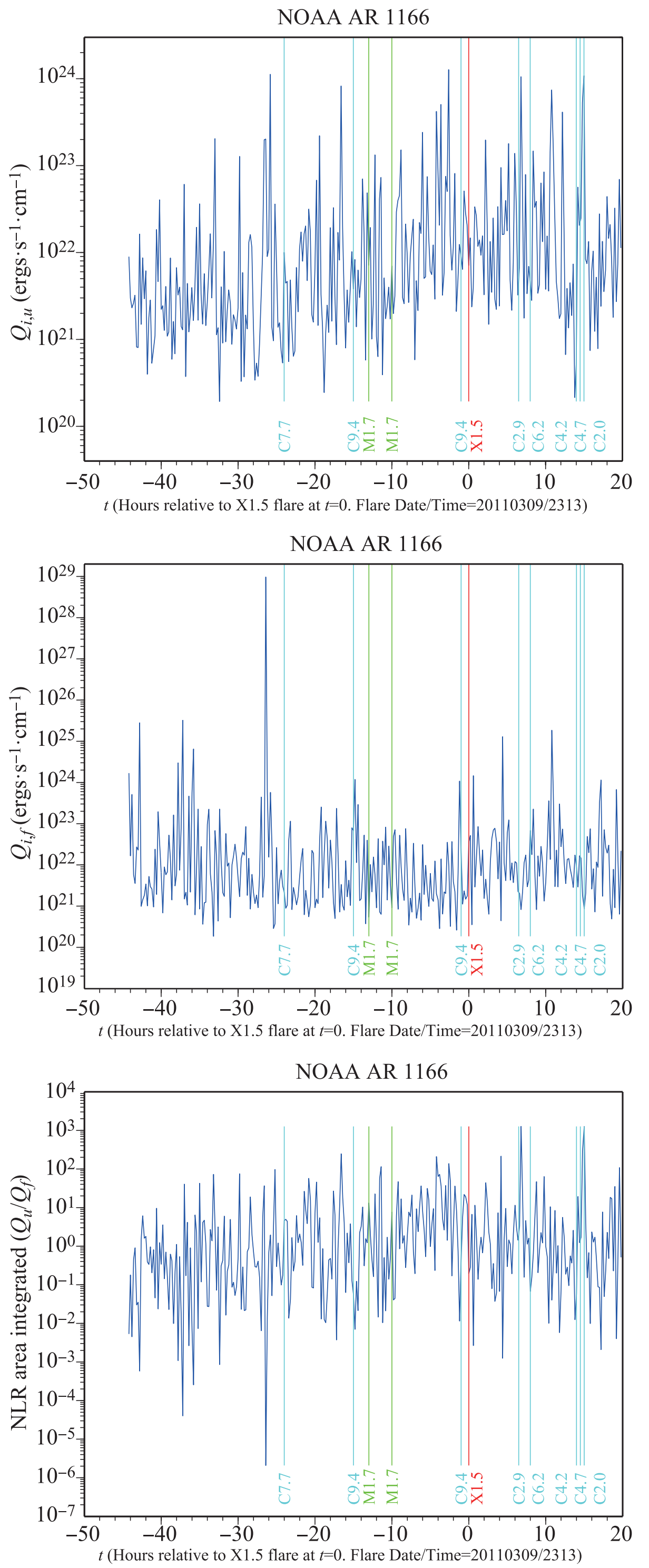

**Fig. 2** Comparison of NLR area integrated filtered and unfiltered heating rates, and their ratio.

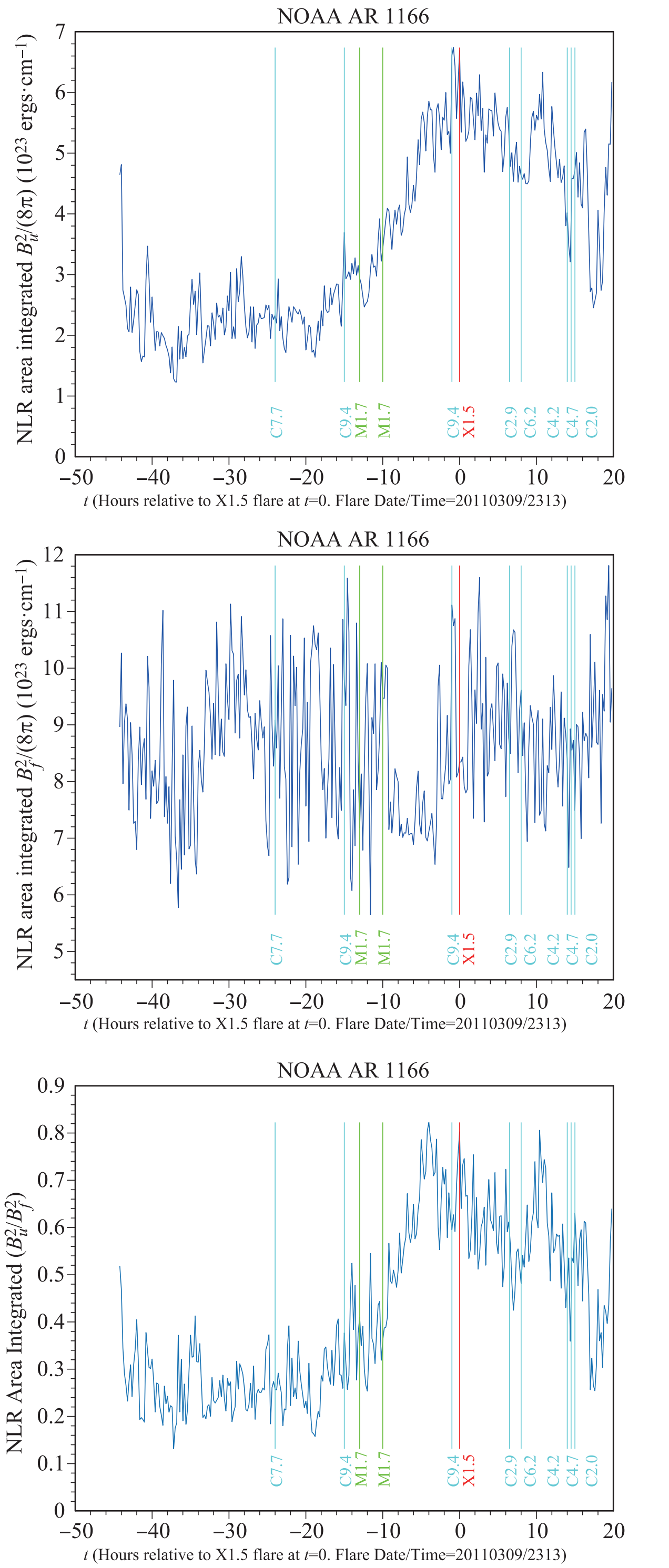

**Fig. 3** Comparison of NLR area integrated filtered and unfiltered magnetic energy density, and their ratio.

$\rho_0\langle v_1^2\rangle + \langle\rho_1(2\boldsymbol{v}_0\cdot\boldsymbol{v}_1 + v_1^2)\rangle$. It is now argued that on the time scale of granular dynamics, the photospheric plasma is incompressible, which here means that $\rho \sim \rho_0$. This is a reasonable approximation if photospheric dynamics is driven mainly by convective forces rather than by magnetic forces. This is the case if two related conditions are satisfied: (i) The first condition is that the plasma $\beta \gg 1$, where $\beta \equiv 8\pi p/B^2 = (4/3)(V_s/V_A)^2$. Here $p, V_s$, and $V_A$ are the pressure, and sound and Alfvén speeds defined by $V_s = [3k_BT/(2m_p)]^{1/2}$ and $V_A = B/(4\pi\rho)^{1/2}$, assuming the ideal gas equation of state $p = \rho k_BT/m_p$ for a neutral H plasma, where $k_B$ and $m_p$ are Boltzmann's constant and the proton mass. Setting $T = 6500$ K, $\rho = 3.2 \times 10^{-7}$ g·cm$^{-3}$, and $B = 500$ G gives $V_s \sim 9$ km·s$^{-1}$, $V_A \sim 2.5$ km·s$^{-1}$, and $\beta \sim 17.3$. Even for field strengths $\sim 10^3$ G, $\beta \sim 4.3$. It follows that photospheric dynamics is dominated by the hydrodynamics of convection, so the speed of acoustic waves determines how rapidly local density fluctuations are damped out by wave spreading. (ii) Given this, the second condition is that $V_s$ be sufficiently large so that the time it takes an acoustic wave to cross a granulation diameter $\sim 10^3$ km is less than the characteristic granulation lifetime $\tau \sim 2.4$–$16.7$ minutes (Section 1.1). This crossing time is $10^3$ km$/V_s \sim 1.85$ minutes, which is $< \tau$. From (i) and (ii) it follows that the effects of $\rho_1$ can be ignored to first order when estimating the fluctuation amplitude $B_1$ within the equi-partition approximation. Then $B_1 \sim (4\pi\rho_0)^{1/2}\langle v_1^2\rangle^{1/2}$. Setting $\langle v_1^2\rangle^{1/2} = 2$–$7$ km·s$^{-1}$, which is the approximate range of granule flow speeds [35] gives $B_1 \sim 400$–$1400$ G, which is $\sim 4$–$14(1.8$–$6.4)$ times larger than the mean(maximum) HMI measurement noise amplitudes for $B$. These rough estimates suggest that the random component of $\boldsymbol{B}$ driven by photospheric dynamics is significantly larger than the HMI measurement error.

The third process is the chaotic component of photospheric dynamics. This is a deterministic component of the dynamics, but usually difficult or impossible to detect in complex, spatially extended systems such as the photosphere due to insufficient resolution, and consequently is often indistinguishable from a random process. This is discussed more in Section 7. Any fluctuations in $\boldsymbol{B}$ due to photospheric chaos is part of the real dynamics of $\boldsymbol{B}$.

The conclusion is that photospheric dynamics dominates HMI measurement error in determining the measured $\boldsymbol{B}$.

In addition, regions of higher $B$ have a higher signal to noise ratio (SNR). NLRs tend to be the regions of highest $B$, and, as discussed in Section 2.3, Schrijver's algorithm [18] constructs NLRs based on selecting $3 \times 3$ pixel blocks containing one or more pairs of opposite polarity LOS magnetic fields with magnitudes $\geq 150$ G, tending to maximize the SNR.

## 5 Heating spikes and signed magnetic flux in strongly flaring ARs

Let $\Phi_i$ be the signed magnetic flux through the photosphere. It is computed as the integral of $B_z$ over the area of the NLR. Figure 4 and the upper left plot in Fig. 5 show the time series of $Q_i$, and of the indicated time averages of $\Phi_i$ for the 7 SF ARs.

### 5.1 A plausible correlation between spikes in $Q_i$, and M and X flares

The times of the largest spikes in $Q_i$ relative to the times of M and X flares in these plots indicate it is plausible that these spikes are correlated with the subsequent occurrence of these flares, and indicate the need to analyze time series of $Q_i$ for more ARs to statistically determine if such a correlation exists. The plausibility of a correlation is suggested by the following trends shown for each AR.

AR 1158: The largest spike in $Q_i$ by a factor $\sim 25$ occurs $\sim 38$–$68$ hours before the first X and M flares, and before the start of the flaring sequence.

AR 1166: The largest spike is $\sim 3.5$ orders of magnitude larger than all others, and occurs 26 hours before the X flare. The next 2 largest spikes occur $\sim 38$–$44$ hours before the X flare. All of these spikes occur before the start of the flaring sequence.

AR 1261: The 6 largest spikes occur $\sim 18$–$39$ hours before the M flare, and all but one are more than an order of magnitude larger than the 7th largest spike. The second largest spike, which is close in magnitude to the largest, occurs before the start of the flaring sequence.

AR 1283: The 3 largest spikes occur $\sim 22$–$25$ hours before the first X flare, and occur before the start of the flaring sequence. The 2 largest of these are more than an order of magnitude larger than the 4th largest spike.

Coupled ARs 1429 and 1430: These ARs are magnetically coupled in the sense that they are merging during the time series. The 9 largest spikes occur about one day before the two X flares near t=0, but after the X1.1 flare. The meaning of the timing of these spikes for ARs 1429/1430, and for AR 1890 is complicated by the fact that the largest spikes occur between X flares.

AR 1890: For values above background values, which are $\lesssim 10^{23}$ ergs·s$^{-1}$·cm$^{-1}$, $Q_i$ increases from the left towards the first X1.1 flare, attains its largest value between the two X1.1 flares, and tends to decrease after the second X1.1 flare.

AR 2017: The largest spike by an order of magnitude occurs $\sim 4$ hours before the X flare. The next 3 largest spikes occur before the start of the flaring sequence, and $\sim 90$–$105$ hours before the X flare.

For the 5 ARs with heating rate time series that begin before the start of the flaring sequence, the largest spike and/or some of the other large spikes occur before the start of the flaring sequence.

### 5.2 A plausible correlation between $|\mathrm{d}\Phi_i/\mathrm{d}t|$, spikes in $Q_i$, and M and X flares

In Figs. 4 and 5 the time series of $\Phi_i$ are averaged using the Matlab moving window function "smooth". Two windows are chosen for the averaging to better reveal trends on relatively short and long time scales. The time series of $\Phi_i$ for the SF ARs suggest the most rapid and prolonged increases or decreases in $\Phi_i$ are correlated with the larger subsequent increases in $Q_i$ within a few hours. A correlation between longer time scale and rapid changes in $\Phi_i$, and the occurrence of subsequent M or X flares is also suggested by the following trends shown for each AR.

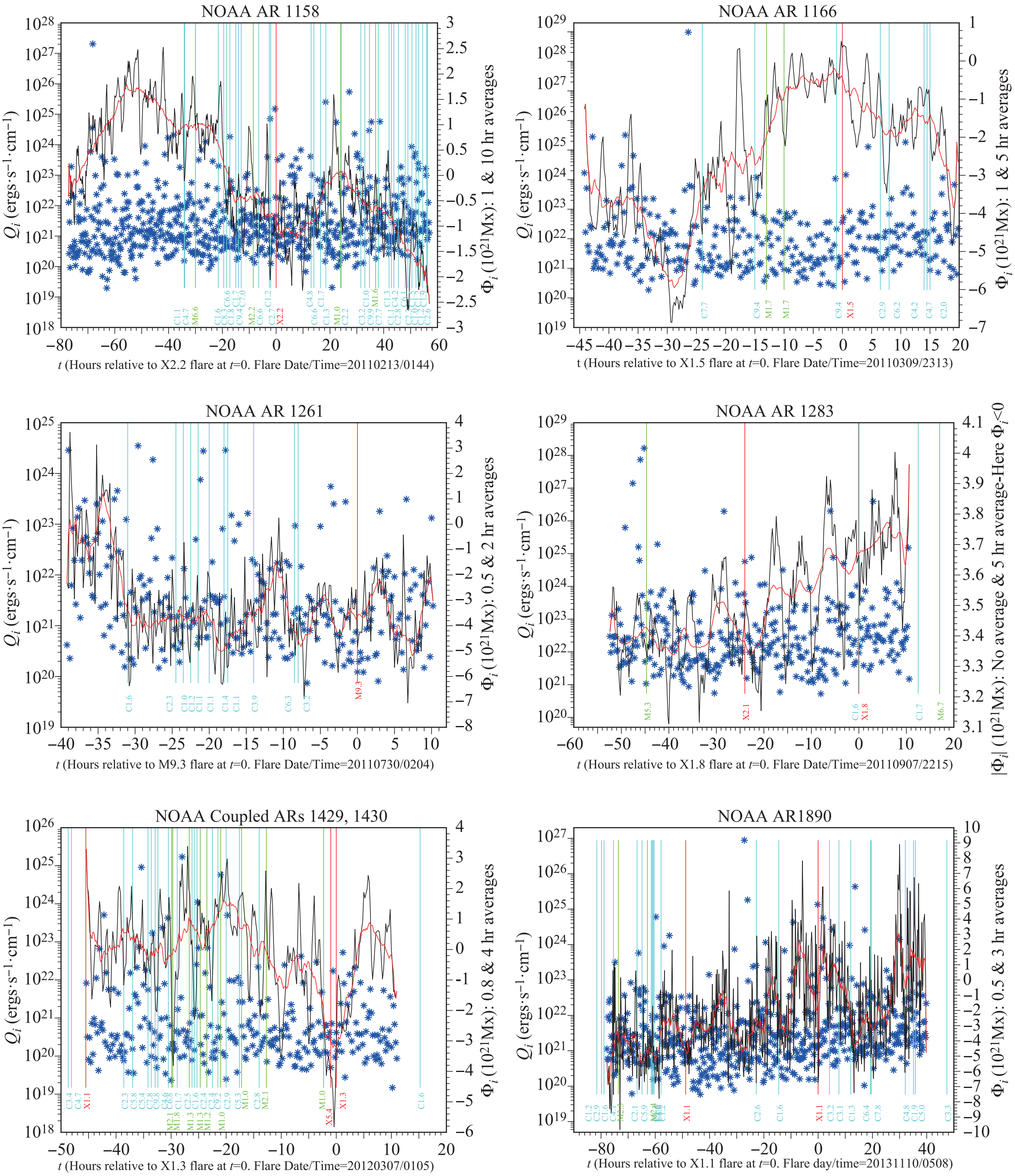

**Fig. 4** $Q_i$ (asterisks), and short (black) and long (red) time averages of $\Phi_i$ for 6 of 7 SF ARs. For AR 1283 the X1.8 and C1.6 flares overlap.

AR 1158: The first M flare occurs about 6 hours after the end of the interval $t = [-55.6, -36.4]$ over which $\Phi_i$ decreases. The second M flare occurs about 6 hours after the end of the interval $t = [-25.4, -14.8]$ over which $\Phi_i$ decreases. The first X flare occurs near the end of the interval $t = [-7.4, 2]$ over which $\Phi_i$ decreases. The third M flare, along with a C flare occurs at the end of the interval $t = [2, 23.6]$ over which $\Phi_i$ increases. $\Phi_i$ then decreases to the end of the time series, with an M flare and a cluster of

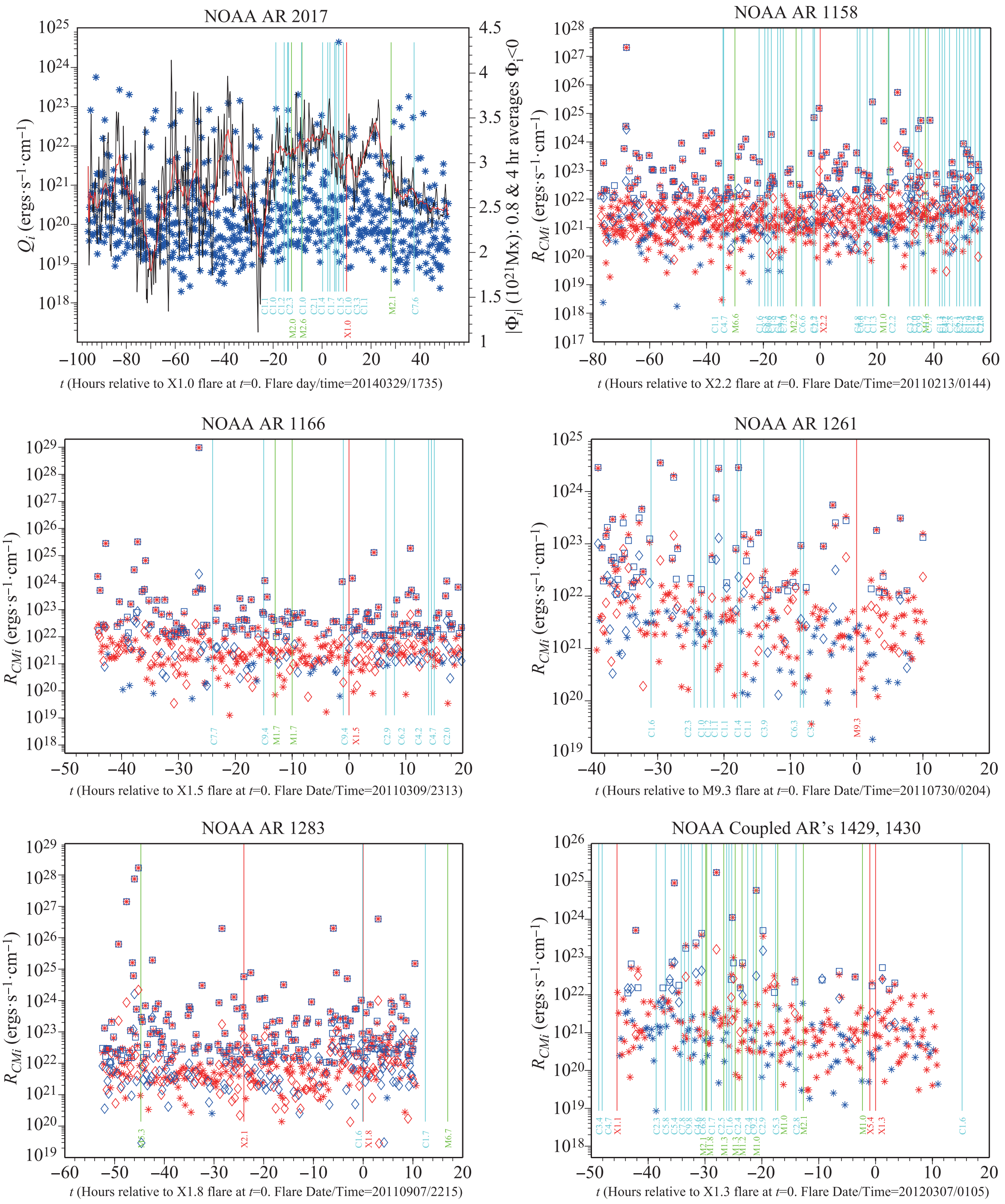

**Fig. 5** $Q_i$, and short (black) and long (red) time averages of $\Phi_i$ for the $7^{th}$ SF AR, and $R_{CMi}$ for 5 of 7 SF ARs. Stars, Squares, and Diamonds label values of $R_{CMi}, Q_i$, and $(\boldsymbol{J} \cdot \boldsymbol{E})_i$. Blue/Red indicates positive/negative values. $Q_i$ and $(\boldsymbol{J} \cdot \boldsymbol{E})_i$ are plotted with $R_{CMi}$ when $Q_i \geq 10^{22}$ ergs·s$^{-1}$·cm$^{-1}$. For AR 1283 the X1.8 and C1.6 flares overlap near $t = 0$.

C flares occurring during this interval. The initial interval from the beginning of the time series to $t = -55.6$, over which $\Phi_i$ increases by a large amount does not contain any flares, but does contain the value of $Q_i$ that is the largest by a factor of 40.

AR 1166: The X flare and the near simultaneous C9.4 flare occur at the end of the 15 hour interval $t = [-45, -30]$, during which $\Phi_i$ decreases. This is followed by the 28 hour interval $t = [-30, -2]$ during which $\Phi_i$ increases, with the two M flares occurring in the middle of this interval after a large amplitude oscillation in $\Phi_i$ in the interval $t = [-20, -16]$. The C7.7 flare at $t = -24$, and the largest value of $Q_i$, at $t = -27$, occur immediately after a period of rapid decrease and increase in $\Phi_i$ over the interval $t = [-35, -27]$.

AR 1261: The M flare at t=0 occurs after a rapid increase and then decrease in $\Phi_i$ during the interval $t = [-18, -8]$. It is less clear if the rapid decrease in $\Phi_i$ in the interval $t = [-35, -30]$, which appears to be associated with at least the first two of the 5 largest values of $Q_i$, is also associated with the M flare.

AR 1283: The first M flare occurs soon after two rapid oscillations in $\Phi_i$, and occurs immediately after the cluster of the 3 largest values of $Q_i$, and after 3 other relatively large values of $Q_i$. The first X flare follows a rapid increase and decrease in $\Phi_i$ over the interval $t = [-31, -25]$. After the first X flare there is an overall increase in $\Phi_i$, beginning with a sharp increase around $t = -20$. There are 3 sharp, large amplitude oscillations in $\Phi_i$ ending $\sim 3$ hours before the second X flare. $\Phi_i$ continues to increase to the end of the time series, with the second M flare occurring $\sim 8$ hours later.

Coupled ARs 1429 and 1430: The time series begins just after the first X flare so there are no prior values of $Q_i$ or $\Phi_i$ that may be correlated with it. The 3 consecutive time intervals defined by the times $t = [-31.2, -26.8, -24.2, -20.2]$ show an overall increase, decrease, and increase, respectively, in $\Phi_i$ with large slopes, and these intervals contain 3, 1, and 2 M flares. There is then an overall decrease in $\Phi_i$ up to $t \sim -10$, with 2 M flares occurring during this interval. After a brief, sharp increase in $\Phi_i$, there is a sharp, large decrease in $\Phi_i$ up to the time of the two X flares near $t = 0$, followed by a sharp, large amplitude increase. The 3 largest values of $Q_i$ occur near the end of intervals during which $\Phi_i$ increases.

AR 1890: The first M flare follows a sharp increase in $\Phi_i$ in the interval $t = [-77, -74]$. The second M flare follows a sharp decrease in $\Phi_i$ in the interval $t = [-67.8, -64.6]$. The first X flare at $t \sim -49$ immediately follows a sharp decrease in $\Phi_i$ beginning at $t \sim -53.2$. The second X flare occurs at the end of the interval $t = [-11.6, 0]$ during which $\Phi_i$ rapidly increases and then decreases. There are other time intervals over which $\Phi_i$ rapidly increases or decreases, and these are often soon followed by large spikes in $Q_i$ or large C flares, or clusters of C flares.

AR 2017: There is a sharp decrease and then increase in $\Phi_i$ over the interval $t = [-50, -30]$, which ends at the beginning of the flaring sequence, and $\sim 10$ hours before the first two M flares. The 4 hour average of $\Phi_i$ then remains fairly constant until it drops sharply over the interval $t = [-9, -1]$, just before the X flare. This is followed by a sharp increase and decrease in $\Phi_i$ just before the subsequent M flare and a cluster of 6 of the largest values of $Q_i$. $\Phi_i$ continues to decrease after this M flare to the end of the time series, with the largest C flare occurring $\sim 10$ hours after the M flare.

A correlation between larger values of $|\mathrm{d}\Phi_i/\mathrm{d}t|$, and subsequent M/X flares and large spikes in $Q_i$ is suggested by the following implications of Faraday's law of induction, $\partial \boldsymbol{B}/\partial t = -c\nabla \times \boldsymbol{E}$. Let $\phi$ be the signed magnetic flux through a photospheric area $A$ bounded by the closed curve $C$. Faraday's law implies $\mathrm{d}\phi(t)/\mathrm{d}t = -c \oint_C \boldsymbol{E} \cdot \mathrm{d}\boldsymbol{l}$, showing that the changing flux generates an induction electric field proportional to $\mathrm{d}\phi/\mathrm{d}t$. Next, taking the curl of Faraday's law, and neglecting the term containing the net charge density, consistent with the quasi-neutrality condition, gives $\partial \boldsymbol{J}/\partial t = [c^2/(4\pi)]\nabla^2 \boldsymbol{E}$, showing that a sufficiently inhomogeneous $\boldsymbol{E}$ drives changes in the current density, increasing (decreasing) the kinetic energy of particles if $\boldsymbol{J} \cdot \boldsymbol{E} > (<)0$. In order for a flare to occur, $\boldsymbol{J} \cdot \boldsymbol{E}$ must be $> 0$ over some volume and time interval in order that magnetic energy be converted into particle energy to drive the flare. These two equations for $\mathrm{d}\phi/\mathrm{d}t$ and $\partial \boldsymbol{J}/\partial t$ may be viewed as applying to each pixel in an NLR. They are applied to an entire NLR by integrating them over the area of the NLR. Let $N(t)$ be the number of pixels in an NLR. Performing this integration on the first equation gives $\mathrm{d}\Phi_i/\mathrm{d}t = -c\Sigma_{j=1}^{N(t)} \oint_{C_j} \boldsymbol{E} \cdot \mathrm{d}\boldsymbol{l} \equiv -c\Sigma_j^{N(t)} V_j(t) \equiv -cV(t)$. Here $V_j(t)$ is the voltage around $C_j$, and $V(t)$ is a total voltage for the NLR. As the net work that would be done by $\boldsymbol{E}$ in moving a unit charge around each of the $C_j$s, $V(t)$ is a measure of the magnetic energy that $\boldsymbol{E}$ can potentially transfer to particles in the NLR at time $t$, so this equation shows a direct connection between $\mathrm{d}\Phi_i/\mathrm{d}t$ and the instantaneous magnetic energy available for conversion into particle energy at the photosphere through the action of the induction electric field.[6)] Taking the scalar product of $\eta\boldsymbol{J}$ with the second equation, for $\partial \boldsymbol{J}/\partial t$, and performing the NLR area integration of the resulting equation, assuming constant $\eta$, implies $\mathrm{d}Q_i/\mathrm{d}t = [c^2\eta/(2\pi)] \int_{NLR} \boldsymbol{J} \cdot \nabla^2 \boldsymbol{E} \mathrm{d}A$. This equation shows a causal connection between the induction electric field, driven by the changing magnetic flux, and the size of $\mathrm{d}Q_i/\mathrm{d}t$, and hence of $Q_i$.

Observations reveal heating events in the photosphere and lower chromosphere with durations $\sim$ 2–5 minutes, and spatial extents $\sim 10^3$ km, which are granulation time and spatial scales, in which plasma is heated to temperatures $\sim 2\times10^4$–$10^5$ K, and accelerated to speeds $\sim$ 50–150 km·s$^{-1}$ [52–62]. The observations of Tian *et al.* [62] show the total AR heating rate due to these events is positively correlated with the magnitude of the emergence rate of the

[6)] If the NLR consists of a single region of connected pixels, $V(t)$ is the voltage around the boundary of this region. If the NLR consists of $n > 1$ disjoint regions, then $V(t) = \Sigma_{k=1}^{n} V_k(t)$, where $V_k(t)$ is the voltage around the boundary of region $k$. Cancellation occurs in this sum if some of the $V_k$ have opposite signs. The NLRs considered here usually consist of more than one region. It is not known if, for $n > 1$, the sum of the magnitudes of the $V_k$ would suggest a stronger correlation with flares and spikes in $Q_i$ than the magnitude of the sum of the $V_k$, which is $|V(t)|(= c^{-1}|\mathrm{d}\Phi_i/\mathrm{d}t|)$. A study that separately analyzes the disjoint components of each NLR is needed to determine this.

total signed magnetic flux for the AR, which corresponds to the $|\mathrm{d}\Phi_i/\mathrm{d}t|$ discussed in this section. Collectively, all of these observations suggest the model predicted spikes in $Q_i$, which as shown in this section are plausibly correlated with $|\mathrm{d}\Phi_i/\mathrm{d}t|$, are examples of these observed heating events.

Time series of the NLR integrated unsigned magnetic flux, computed as $\int_{NLR} |\phi(x, y, 0)|\mathrm{d}A$, and of the magnitude of the NLR integrated signed magnetic flux, denoted $|\Phi_i|$, do not suggest a correlation with $Q_i$ or flares. This is because, in contrast to the signed magnetic flux $\Phi_i$, they exclude negative values, and so do not provide as accurate a representation of the time derivative of the flux, and hence of the effect of the induction electric field, as $\Phi_i$.

### 5.3 The lack of correlation between photospheric magnetic energy and spikes in $Q_i$, and M and X flares

Analysis of active region magnetograms generated from Michelson Doppler Imager (MDI) data shows a lack of correlation between photospheric magnetic energy and coronal flares (e.g., [15]). Consistent with this finding, and as discussed in more detail in the following paragraph, there is no correlation between photospheric magnetic energy and spikes in $Q_i$, or M and X flares for the ARs analyzed here. Three possible reasons for this lack of correlation are given.

Time series of the NLR area integral of the magnetic energy density, denoted $(B^2/(8\pi))_i$, were also plotted along with $Q_i$ and flare times. The plots are not included here. For some SF ARs they show some combination of a gradual increase, or relatively large spikes in $(B^2/(8\pi))_i$ before the beginning of a flaring sequence, or before M or X flares during a flaring sequence. For the other SF ARs, such a correlation between $(B^2/(8\pi))_i$, $Q_i$, and flare times is much less apparent, and this is more the case for the C ARs. Therefore $(B^2/(8\pi))_i$ is much less plausibly correlated than $|\mathrm{d}\Phi_i/\mathrm{d}t|$ with the larger spikes in $Q_i$, and M and X flares. Three possible reasons for the lack of such a correlation are the following. First, $(B^2/(8\pi))_i$ includes potential and non-potential magnetic energy, and so might not be an accurate measure of the non-potential energy, which is the free magnetic energy available for conversion into particle energy. Second, $(B^2/(8\pi))_i$ is a photospheric quantity, and so might not provide any information about the magnetic energy in the corona, where the flaring process is believed to occur. The third reason is based on the time evolution equation for $B^2/(8\pi)$, which is Poynting's theorem, given by $(8\pi)^{-1}\partial B^2/\partial t = -\nabla\cdot\boldsymbol{S} - \boldsymbol{J}\cdot\boldsymbol{E}$, where $\boldsymbol{S}$ is the Poynting flux. Integrating this over an NLR gives $\mathrm{d}(B^2/8\pi)_i/\mathrm{d}t = -\int_{NLR}\nabla\cdot\boldsymbol{S}\mathrm{d}A - (Q_i + R_{CMi})$. As discussed in Section 5.4, the largest spikes in $Q_i$ are found to be driven by the conversion of CM KE into thermal energy, rather than by the conversion of magnetic energy into thermal energy, meaning that $Q_i \sim -R_{CMi}$ during a large spike. In that case it follows from Poynting's theorem that during a large spike any change in $(B^2/(8\pi))_i$ is mainly due to the Poynting flux term, and so does not have a significant correlation with $Q_i$.

### 5.4 The large spikes in $Q_i$ are due to non-force-free currents and convection driven heating

Figures 5 and 6 show the time series of $R_{CMi}$ for the 7 SF ARs. The plots include the values of $Q_i$ and $(\boldsymbol{J}\cdot\boldsymbol{E})_i$ when $Q_i$ exceeds a threshold for each time series. Each threshold is the approximate value below which most values of $Q_i$ occur, and above which the relatively few large spikes occur. The time series show that for values of $Q_i$ above the threshold, which is similar ($\sim 10^{21}$–$10^{22}$ ergs·s$^{-1}$·cm$^{-1}$) for all 7 time series, $Q_i$ is very close to $-R_{CMi}$. As discussed in Section 2.5, this is the case of convection driven heating: the largest heating events, corresponding to the largest current spikes, are due to the conversion of bulk flow kinetic energy into thermal energy, rather than due to the conversion of magnetic energy into thermal energy. If convection driven heating occurs in this way, it does not exclude the possibility that these heating events are reconnection driven, with magnetic energy initially converted into bulk flow kinetic energy in the form of reconnection jets with a component $\perp\boldsymbol{B}$ that generates a convection electric field $(\boldsymbol{v}\times\boldsymbol{B})/c$ that drives a $J_\perp$ that undergoes resistive dissipation. This possibility corresponds to the case for which the space and time averaged $\boldsymbol{B}$ inferred by HMI mainly describes the latter part of this process.

Let $I_\perp$ and $I_{//}$ be the NLR integrals of $J_\perp$ and $J_{//}$. $I_\perp$ and $I_{//}$ are the unsigned NLR currents. Then $Q_i = \eta(I_\perp^2 + I_{//}^2)$. Figures 6–8 compare $I_\perp$ and $I_{//}$ for each of the 14 ARs. Figures 6 and 7 show that for the SF ARs, $I_\perp > I_{//}$ at almost all times, that $I_\perp$ is orders of magnitude larger than $I_{//}$ at all times of moderate or larger increases in $I_\perp$, and that $I_{//}$ is essentially constant, or varies by no more than a factor $\sim$ 2–3 for all times. Since $Q_i \propto$ the squares of the currents, it follows that $Q_i \sim \eta I_\perp^2$ at almost all times. Resistive heating in the SF ARs is due almost entirely to dissipation of non-force-free currents. For the C ARs, Figs. 7 and 8 show that $I_\perp > I_{//}$ most of the time, and that $I_\perp^2$ is orders of magnitude larger than $I_{//}^2$ for moderate or larger increases in $I_\perp$. For these increases, $Q_i \sim \eta I_\perp^2$. Even in the C ARs, resistive heating by non-force-free currents tends to be dominant.

## 6 Heating spikes in control ARs

Figure 9 shows plots of $R_{CMi}$ for 6 of 7 C ARs. The plots include the values of $Q_i$ and $(\boldsymbol{J}\cdot\boldsymbol{E})_i$ when $Q_i$ exceeds a threshold, chosen in the same way as for Figs. 5 and 6. Similar to the SF ARs, the larger heating rates are convection driven.

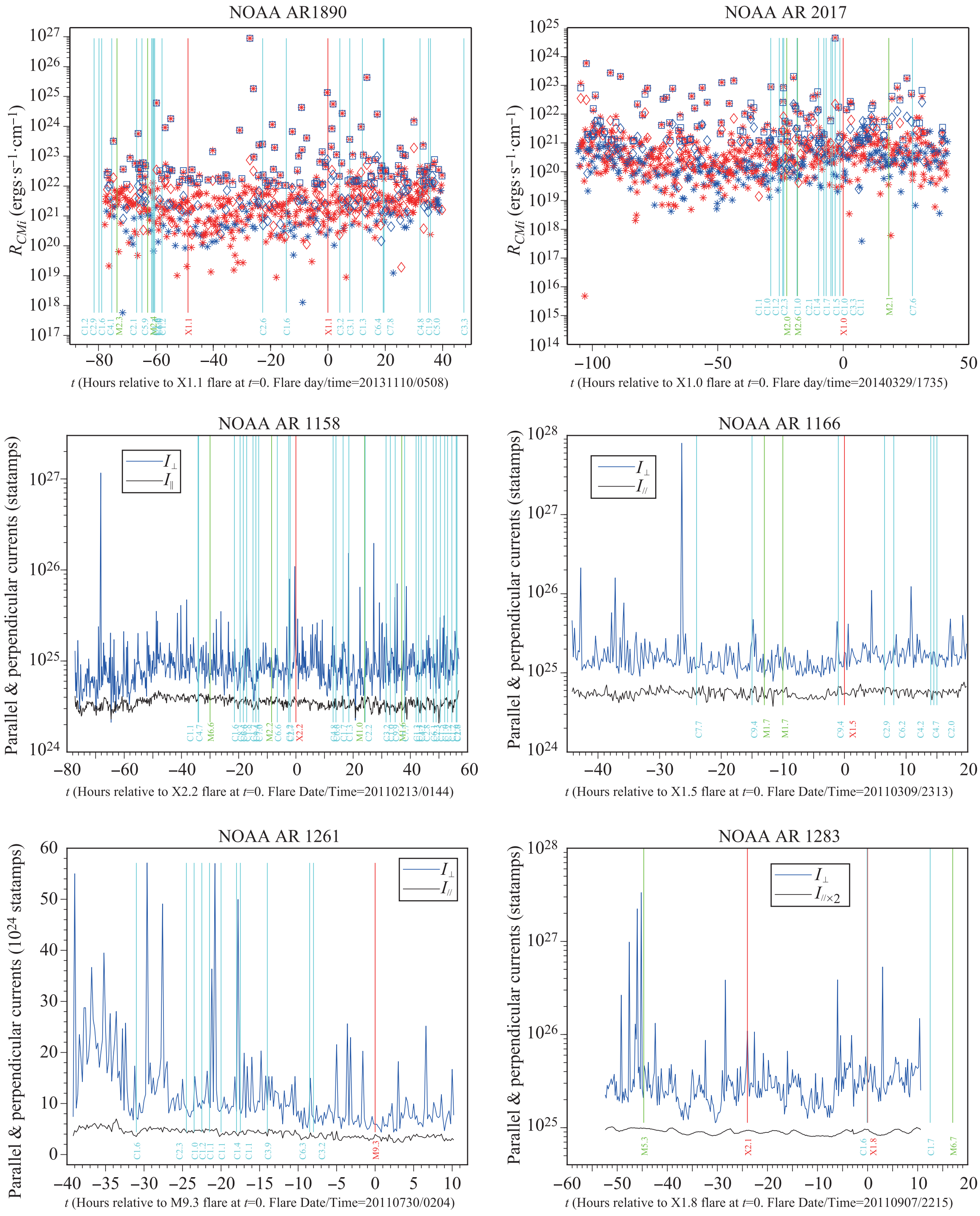

**Fig. 6** $R_{CMi}$ for the remaining 2 of 7 SF ARs, and $I_\perp$ and $I_{//}$ for 4 of 7 SF ARs. Stars, Squares, and Diamonds label values of $R_{CMi}, Q_i$, and $(\boldsymbol{J}\cdot\boldsymbol{E})_i$. Blue/Red indicates positive/negative values. $Q_i$ and $(\boldsymbol{J}\cdot\boldsymbol{E})_i$ are plotted when $Q_i \geq 10^{22}(10^{21})$ ergs·s$^{-1}$·cm$^{-1}$ for the left (right) plot.

Figures 5, 6, and 9 show that the $Q_i$ values that clearly indicate an increase in heating by orders of magnitude over background values are $\sim 10^{24}$–$10^{29}$ ergs·s$^{-1}$·cm$^{-1}$ for the SF ARs, and $\sim 10^{22}$–$10^{24}$ ergs·s$^{-1}$·cm$^{-1}$ for the C ARs. This suggests that the occurrence of $Q_i$ values in the first range in an AR is plausibly predictive of the subsequent occurrence of M or X flares, while the occurrence of $Q_i$ values in the second range in an AR is plausibly predictive of flares not larger than C class. However, comparison of the figures also suggests that a correlation of $Q_i$ values in

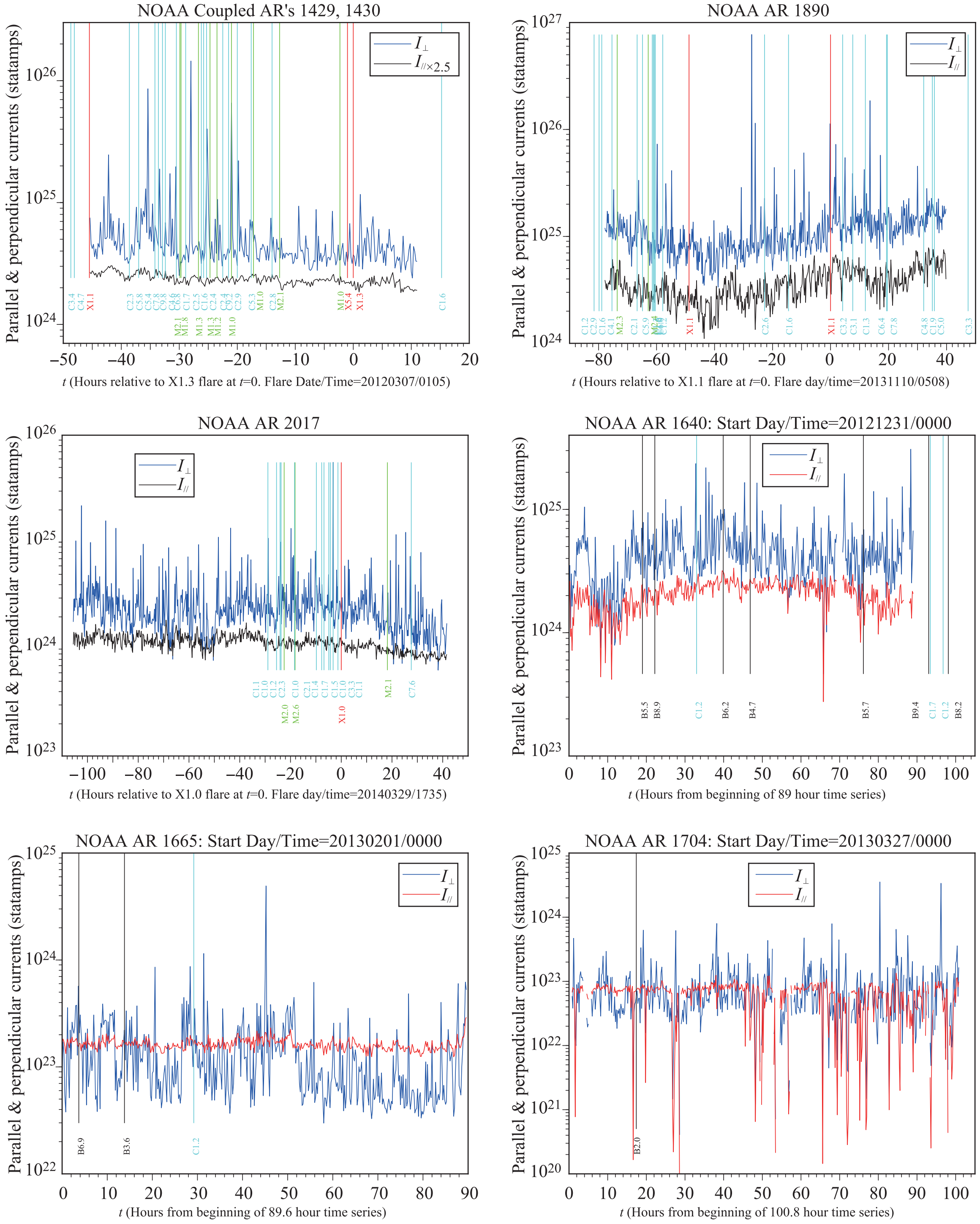

**Fig. 7** $I_\perp$ and $I_{//}$ for the remaining 3 of 7 SF ARs (1429/1430, 1890, 2017), and for 3 of 7 C ARs (1640, 1665, 1704).

the second range with the subsequent occurrence of C or weaker class flares is less plausible than a correlation of $Q_i$ values in the first range with the subsequent occurrence of M and X flares.

For all 14 ARs, the subset of the largest spikes analyzed at the pixel level are found to occur on the HMI and granulation scales of 1 arcsec and 12 minutes.[7)] Spikes are

[7)] Suppression of convection in high magnetic field regions increases the convection time scale above this characteristic 12 minute value. This makes the ratio of the HMI temporal resolution to the convection time scale smaller, so regions with suppressed convection are temporally resolved to a higher degree.

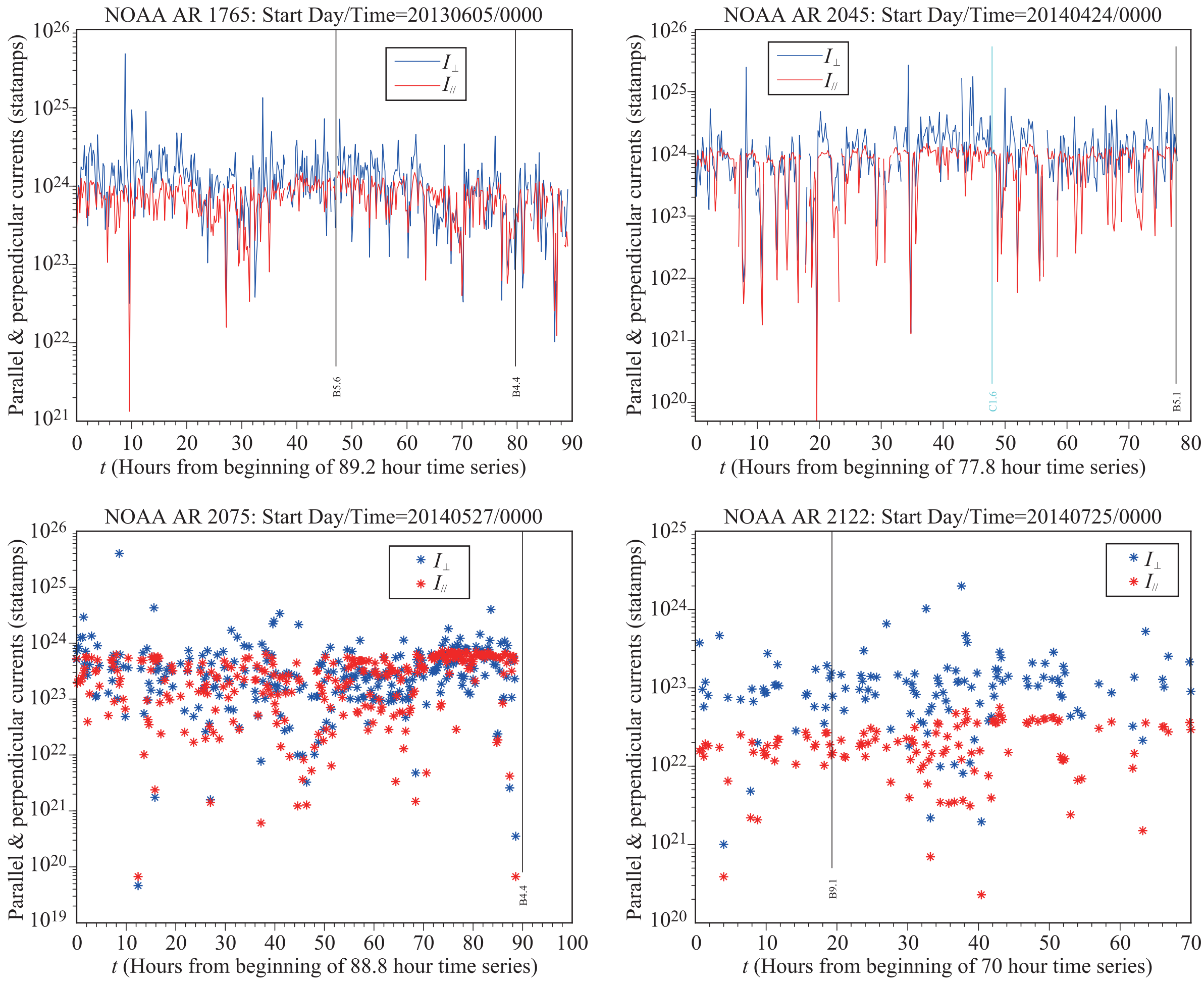

**Fig. 8** $I_\perp$ and $I_{//}$ for the remaining 4 of 7 C ARs. The data labeling is changed from line to asterisk for ARs 2075 and 2122 because for these ARs the NLR does not exist for so many times that the line plot function fails to interpolate across all data gaps. As discussed in Section 2.3, in order for a pixel block to be included in the NLR it must contain one or more pairs of opposite polarity LOS magnetic fields with magnitudes > 150 G. At certain times, especially for the C ARs, which tend to have weaker fields, there may be no pixels for which this threshold is exceeded, in which case there is no NLR.

found in ARs with and without M or X flares, and outside as well as inside NLRs, but the largest spikes are localized in the NLRs of ARs with M or X flares.

## 7 Similarity of the CDFs of $Q_i$ and coronal flares, and the possibility of self organized criticality in the photospheric heating rate distribution

The CDFs of $Q_i$ for each of the 14 ARs, for all SF ARs collectively, for all C ARs collectively, and for all ARs collectively are shown in Figs. 10–14. Log–log plots are used since if $N(Q_i) = AQ_i^{-s}$ where $A$ and $s$ are constant over some range of $Q_i$, then $\log N(Q_i) = -s \log Q_i + \log A$ over this range, which is a line with slope $-s$. CDFs with these properties are called scale invariant since the value of $s$ does not depend on any physical scale over this range of $Q_i$. The lines that are fit to the data are generated by the Matlab polyfit function over the range of $Q_i$ indicated in each figure. In most cases the linear fits are very good over the indicated ranges. In some cases the linear behavior breaks down for $N(Q_i) \lesssim 10$. Visual inspection suggests that in most cases this breakdown is due to an insufficient number of data points for the higher values of $Q_i$, in which case the linear behavior would extend to higher values of $Q_i$ if longer time series were used to provide better statistics. However, for SF AR 1166 in Fig. 10 there is a data point at $Q_i = 10^{29}$ ergs·s$^{-1}$·cm$^{-1}$, about 3.5 orders of magnitude to the right of the bottom of the linear fit. This suggests that for this AR the scale invariant behavior breaks down for $Q_i \gtrsim 10^{26}$ ergs·s$^{-1}$·cm$^{-1}$.

The figures show that the CDFs of the 14 ARs exhibit scale invariant behavior over ranges of $Q_i$ that extend over $\sim$ 2–2.9 orders of magnitude for 3 ARs, and over $\sim$ 3.3–6.3 orders of magnitude for the remaining 11 ARs. The CDFs of all ARs, all SF ARs, and all C ARs show scale invariant behavior over a range of $Q_i$ that extends over $\sim$ 4–5 orders of magnitude. This behavior is evidence that whatever process generates $Q_i$ above an AR dependent threshold value remains the same over the corresponding range of $Q_i$. This behavior is a necessary but not sufficient condition for the system, here an AR NLR, to be in a state of self organized criticality (SOC) [63]. SOC states arise naturally during the evolution of dynamical systems with extended spatial degrees of freedom, such as the photosphere and corona. They are characterized by long range spatial order, are stable in a narrow range of system parameter values, and exhibit a high degree of chaos and noise [63–69], where chaos is deterministic but often indistinguishable from noise due to insufficient resolution. The transition of a system into an SOC state is similar to a phase transition in that the system makes a transition from a state without long range order to one with long range order.

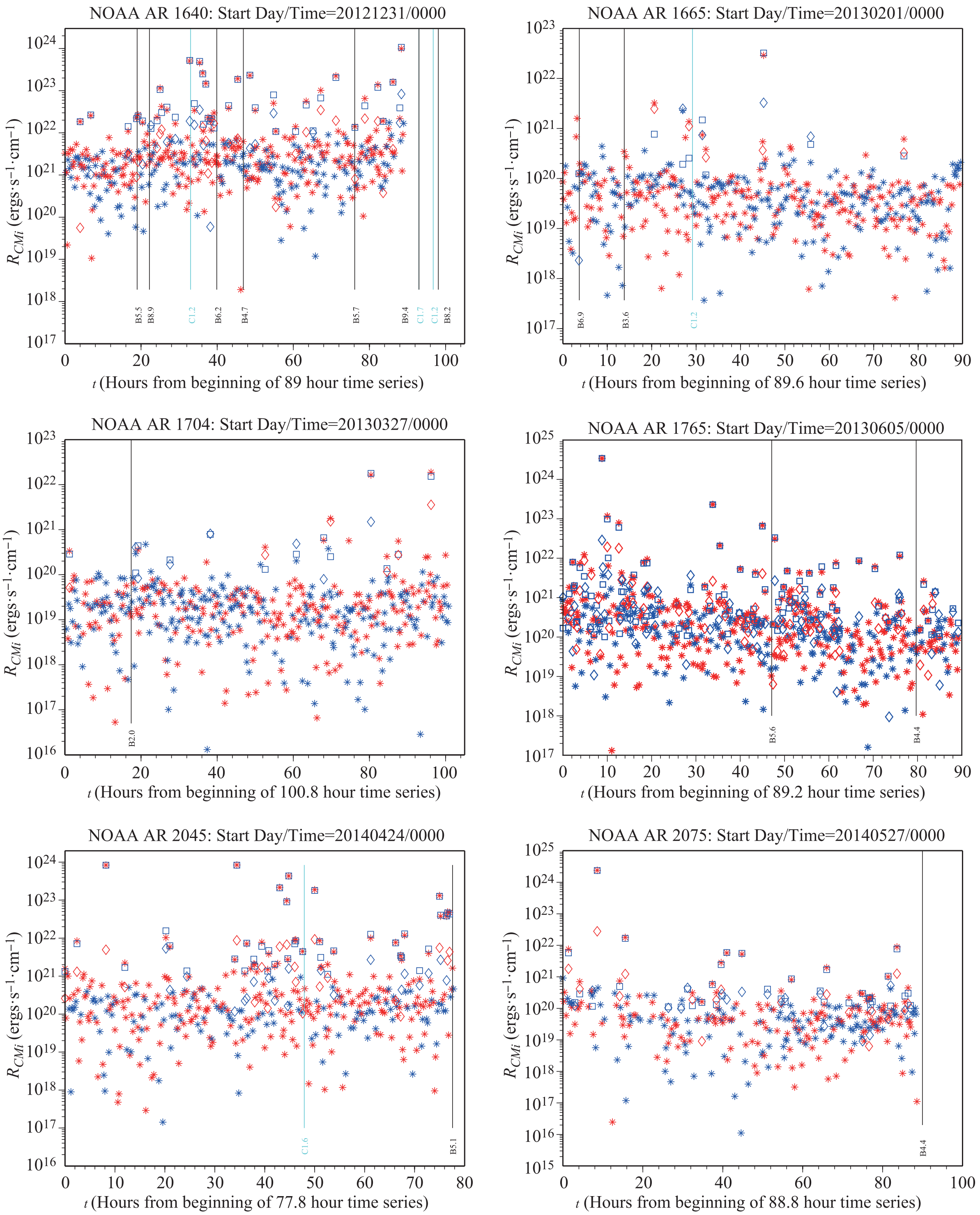

**Fig. 9** $R_{CMi}$ for 6 of 7 C ARs. The plot for the $7^{th}$ C AR is similar to these. Stars, Squares, and Diamonds label values of $R_{CMi}, Q_i$, and $(\boldsymbol{J} \cdot \boldsymbol{E})_i$. Blue/Red indicates positive/negative values. $Q_i$ and $(\boldsymbol{J} \cdot \boldsymbol{E})_i$ are plotted when $Q_i \geq 10^{20}, 10^{21}$, or $10^{22}$ ergs·s$^{-1}$·cm$^{-1}$ depending on the plot. B flares are included, labeled by black lines.

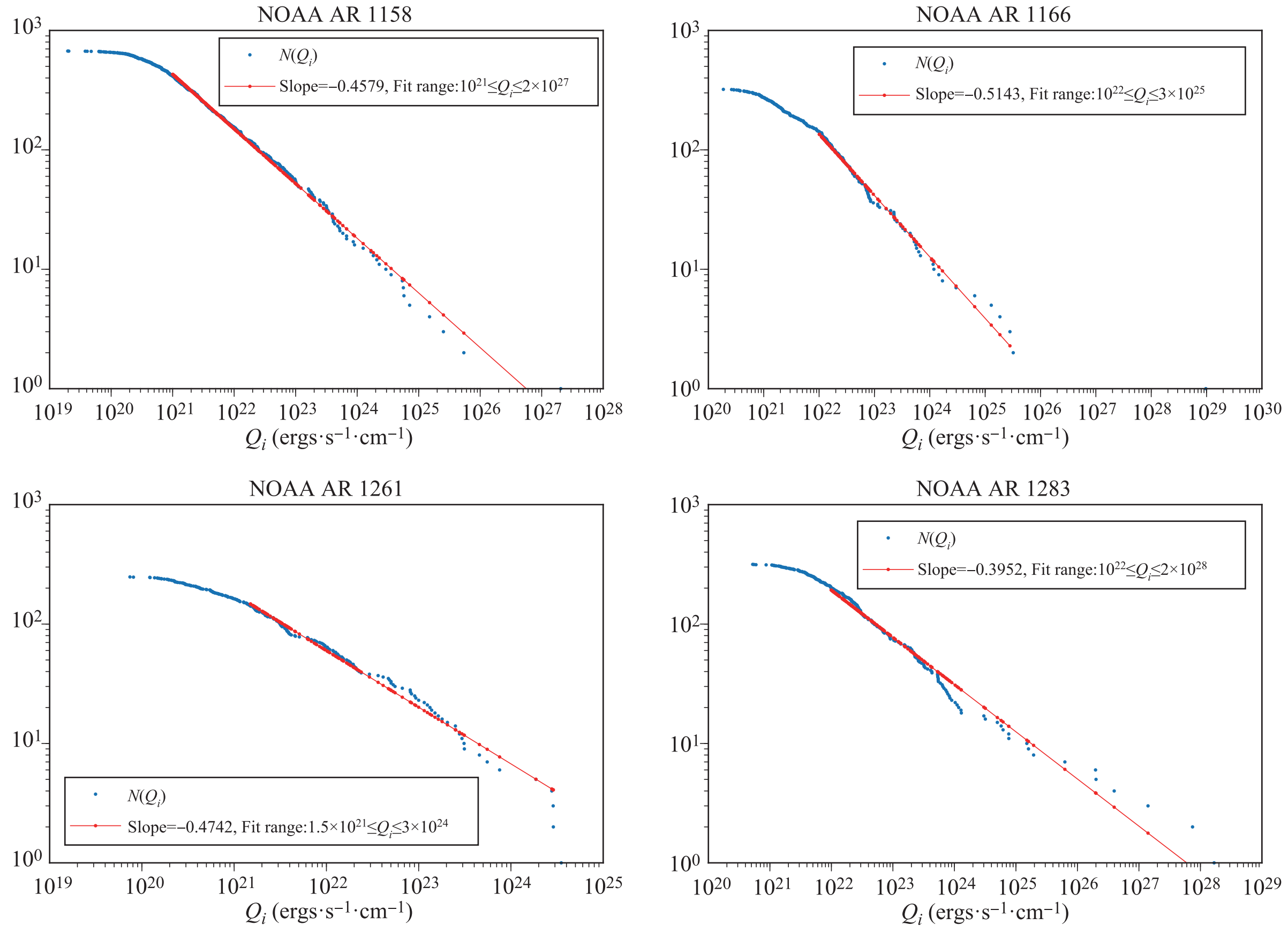

**Fig. 10** CDFs of $Q_i$ for SF ARs 1158, 1166, 1261, and 1283. The number of orders of magnitude over which the linear fits extend are respectively 6.3, 3.48, 3.3, and 6.3.

For the SF ARs, $s$ = (0.4472, 0.4579, 0.4708, 0.4742, 0.4858, 0.5101, 0.5143). For the C ARs, $s$ = (0.3351, 0.4080, 0.4303, 0.4356, 0.4459, 0.5148, 0.5385). The range of $s$ is $\sim$ 0.34–0.54, with all but one value in the range 0.4–0.54. The mean, median, and standard deviation of $s$ for all SF ARs, all C ARs, and all ARs are, respectively, $(0.4800, 0.4742, 0.0252)$, $(0.4440, 0.4356, 0.0675)$, and $(0.4620, 0.4643, 0.0524)$. This shows there is little statistical variation in $s$ among the 14 ARs, so $s$ is largely independent of individual AR properties.

The scale invariant behavior of $N(Q_i)$, and the range, mean values, and weak dependence of $s$ on AR are similar to the observed properties of the CDF $N(E)$ for coronal flares, where $E$ is the total energy released, defined as the total amount of magnetic energy converted into particle energy. It is observed that $N(E) \propto E^{-\alpha_E}$ where $\alpha_E$ varies little from one AR to another, being largely independent of the differences between ARs such as area, total unsigned magnetic flux, and sunspot distribution [70–72]. Observation based estimates of $\alpha_E$ have been made for over 40 years [70, 73–79]. For HXR, SXR, and $\gamma$ ray based constructions of $N(E)$, it is found that $\alpha_E \sim 0.40$–$0.88$, with almost all values in the range of $\sim 0.4$–$0.6$.

These observation based values for $\alpha_E$ are consistent with: (i) Values of $\alpha_E$ predicted by first principles, SOC theories, which give $\alpha_E \sim 0.4$–$0.67$ [76–79]. (ii) Values of $\alpha_E$ predicted by simulations of the solutions to lattice based avalanche models, which give $\alpha_E \sim 0.40$–$0.57$ [76, 77, 80–83]. (iii) Values of $\alpha_E \sim 0.35$–$0.6$ predicted by semi-empirical, force-free magnetic field models for the free energy distribution of magnetic discontinuities in non-flaring AR coronae [84].

The fact that the ranges of $s$ and $\alpha_E$ are similar, and that $s$ and $\alpha_E$ show relatively little variation with AR suggests a common origin of the photospheric $N(Q_i)$ and the coronal $N(E)$. If they do have a common origin, a question is whether SOC theory can help clarify it. The first application of SOC theory to propose a physical process that gives rise to $N(E)$ is presented in Refs. [80, 81]. There it is proposed that the solar corona is in an SOC state, and that flares consist of a time series of random, spatially distributed avalanches of sub-resolution magnetic reconnection events that trigger one another. If this theory of flares as a coronal process is correct, a question is whether the photosphere is also in an SOC state, and exhibits similar avalanche type flaring events with a similar CDF on smaller energy and spatial scales. The form of $N(Q_i)$ found here for 14 ARs suggests the answer to this question may be affirmative for the SF ARs, which have the largest $Q_i$ spikes. For C ARs, the results presented here suggest their photospheres are in an SOC state, but the connection between this apparent property and flaring in these regions is less evident. The similarity of $N(E)$ and $N(Q_i)$ discussed above does not prove ARs enter an SOC state when $Q_i$ exceeds a threshold value, but it is evidence in support of this possibility, and implies it is important to further explore this possibility using time series from more ARs for better statistics.

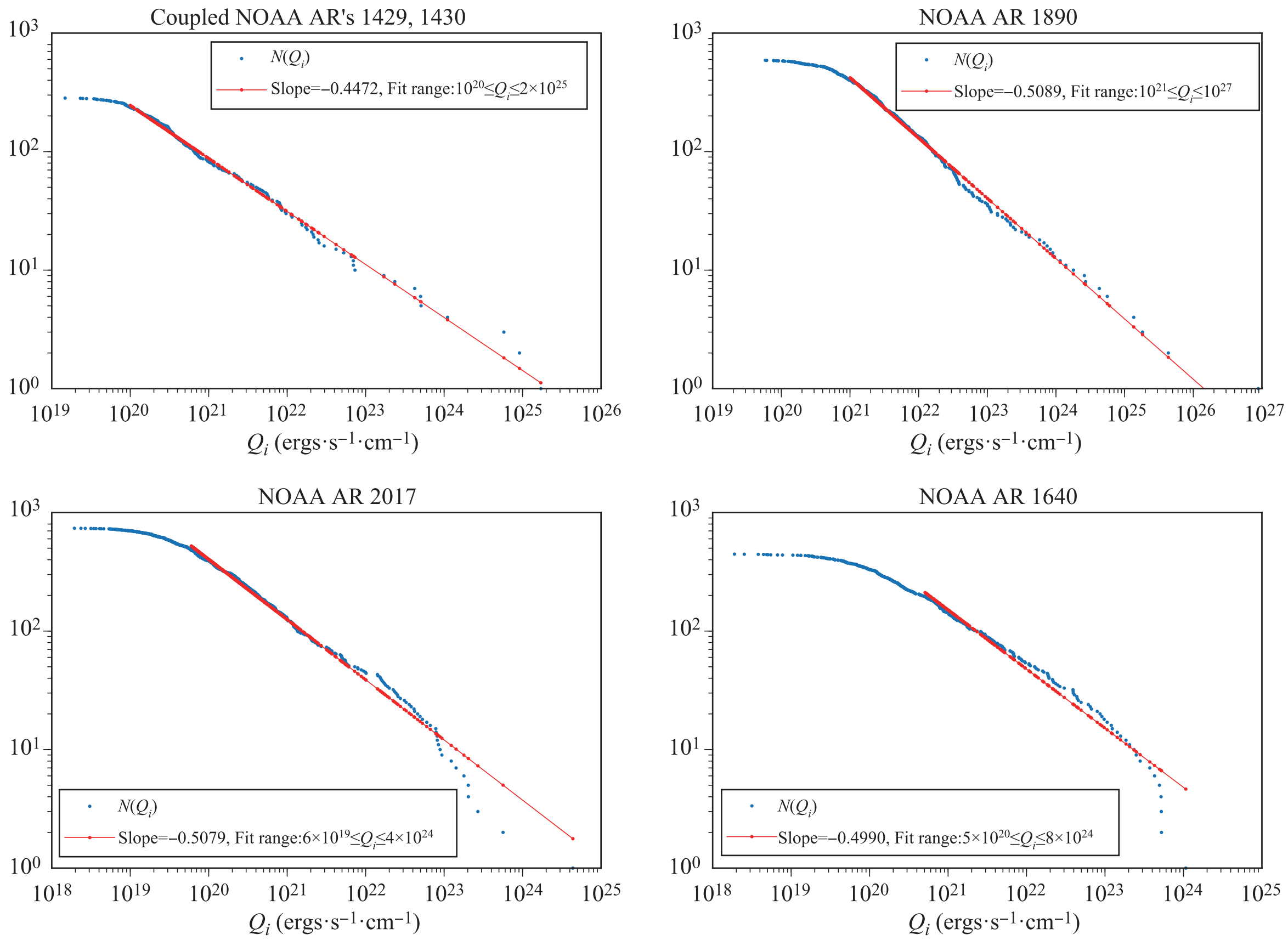

**Fig. 11** CDFs of $Q_i$ for SF ARs 1429/1430 (coupled), 1890, and 2017, and C AR 1640. The number of orders of magnitude over which the linear fits extend are respectively 5.3, 5.82, 4.82, and 4.2.

SOC related reconnection avalanches are not the only possible process that can give rise to the observed $N(E)$. A model based on the idea of a flare being due to a single energy release event with a range of possible energies, and of the observed $N(E)$ being due to the statistical distribution of such events is presented by Wheatland & Craig [85].

The similarity of $N(Q_i)$ and $N(E)$, and the suggestion that photospheric AR NLRs are in an SOC state are consistent with the observation based findings of Uritsky *et al.* [86, 87] for the CDFs of photospheric magnetic field fluctuations and overlying coronal heating events in the quiet Sun. Uritsky *et al.* [86] generate time series of increases in the MDI LOS magnetic flux, and in the intensity of overlying coronal EUV emission flux images from the Solar Terrestrial Relations Observatory Extreme Ultraviolet Imager (STEREO EUVI, [88]) in quiet Sun regions. Define the CDF $P(I)$ of each of these time series as the probability of the occurrence of an intensity or LOS magnetic flux increase $\geq I$. Uritsky *et al.* [86] computed $\mathrm{d}P(I)/\mathrm{d}I$ for each time series, and show they are scale invariant power law distributions with exponents $1.48\pm0.03$ for the photospheric LOS magnetic flux, and $1.47\pm0.03$ for the coronal EUV emission flux (see table 3 of Ref. [86]). It follows that the corresponding $P(I)$, which are the integrals of $\mathrm{d}P(I)/\mathrm{d}I$ over $I$, are also scale invariant power law distributions, with exponents $\sim 0.48$ and 0.47, respectively, which are essentially identical. Further analysis by Uritsky

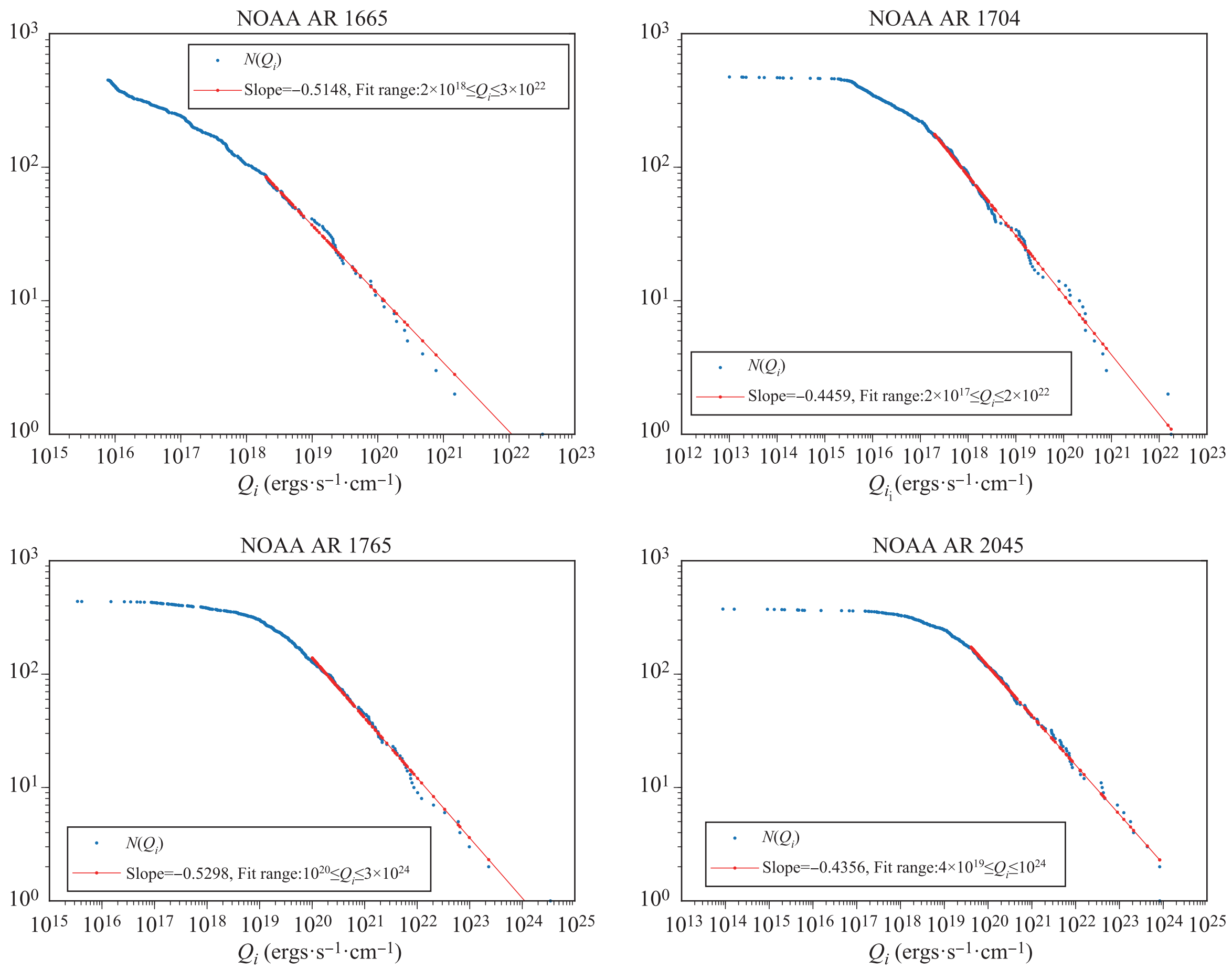

**Fig. 12** CDFs of $Q_i$ for C ARs 1665, 1704, 1765, and 2045. The number of orders of magnitude over which the linear fits extend are respectively 4.18, 5, 4.48, and 4.4.

*et al.* [87] leads to the conclusion that in the quite Sun, the photosphere and corona are in magnetically coupled SOC states, the quiet corona is heated by SOC-like avalanches of magnetic energy dissipation, and the heating on spatial scales $> 3 \times 10^3$ km is driven by turbulent photospheric convection. Uritsky *et al.* [86] identified increases in the photospheric LOS magnetic flux as proxies for energy release events in the photosphere, and show they are correlated with coronal heating events identified by increases in EUVI flux. This raises the question of whether increases in the AR NLR heating rate $Q_i$ computed here is correlated with the overlying coronal heating rate. Another study involving coronal observations is necessary to answer this question, and might also give insight into a possible correlation between the times of enhancements in $Q_i$, and the times of subsequent flaring.

## 8 180 degree ambiguity error

There is 180 degree ambiguity phase error in the HMI data [34]. It results from the fact that HMI can determine the direction of the magnetic field component perpendicular to the LOS, denoted $\boldsymbol{B}_{\perp LOS}$, only up to a sign, so the direction can be in error by an angle $\pi$. The error tends to be larger (smaller) in regions where $B$ is smaller (larger). The error can cause a spurious gradient in $\boldsymbol{B}_{\perp LOS}$ between adjacent pixels, and a corresponding change in $\boldsymbol{J}$. Since many of the results in this paper are based on a calculation of $\boldsymbol{J}$, it is important to estimate if ambiguity phase error causes significant error in this calculation.

Phase error correction algorithms are implemented in the HMI data pipeline by the HMI team. The algorithms are not perfect, so it is important to estimate the residual ambiguity error. In the Cartesian coordinate system used here, this is done using two tests, one global, and one local. Both tests are based on using time sequences of images of the horizontal magnetic field $\boldsymbol{B}_h = (B_x, B_y)$ in an AR to determine where $\boldsymbol{B}_h$ reverses direction in one time step. This is done for several ARs. Here $\boldsymbol{B}_h \sim \boldsymbol{B}_{\perp LOS}$ for the near-disk centered ARs used here. It is assumed that if $\boldsymbol{B}_h$ changes direction by $\pm\pi$ to within 1% in a pixel in one time step, that change is due to ambiguity error, rather than real magnetic field dynamics.

The global test consists of randomly selecting 5 time consecutive images of $B_x$ and $B_y$ in each AR pixel, com-puting the angle $\theta = \arctan(B_y/B_x)$ that $\boldsymbol{B}_h$ makes with the $x$ axis in each pixel for each image, and then using these 5 global images of $\theta$ to compute the 4 time consecutive difference images of the change $\Delta\theta$ in $\theta$ in each pixel. From each of these 4 images, an image is then created labeling only those pixels for which $\Delta\theta$ is in the range $0.99\pi \leqslant |\Delta\theta| \leqslant 1.01\pi$. These are the pixels with ambiguity error. In all cases it is found that the NLRs, where $B$ tends to be largest, are almost empty of such pixels, but there are a significant number of such pixels outside the NLRs, where $B$ tends to be smaller by a factor $\sim 10$. This test covers a time interval of 1 hour.

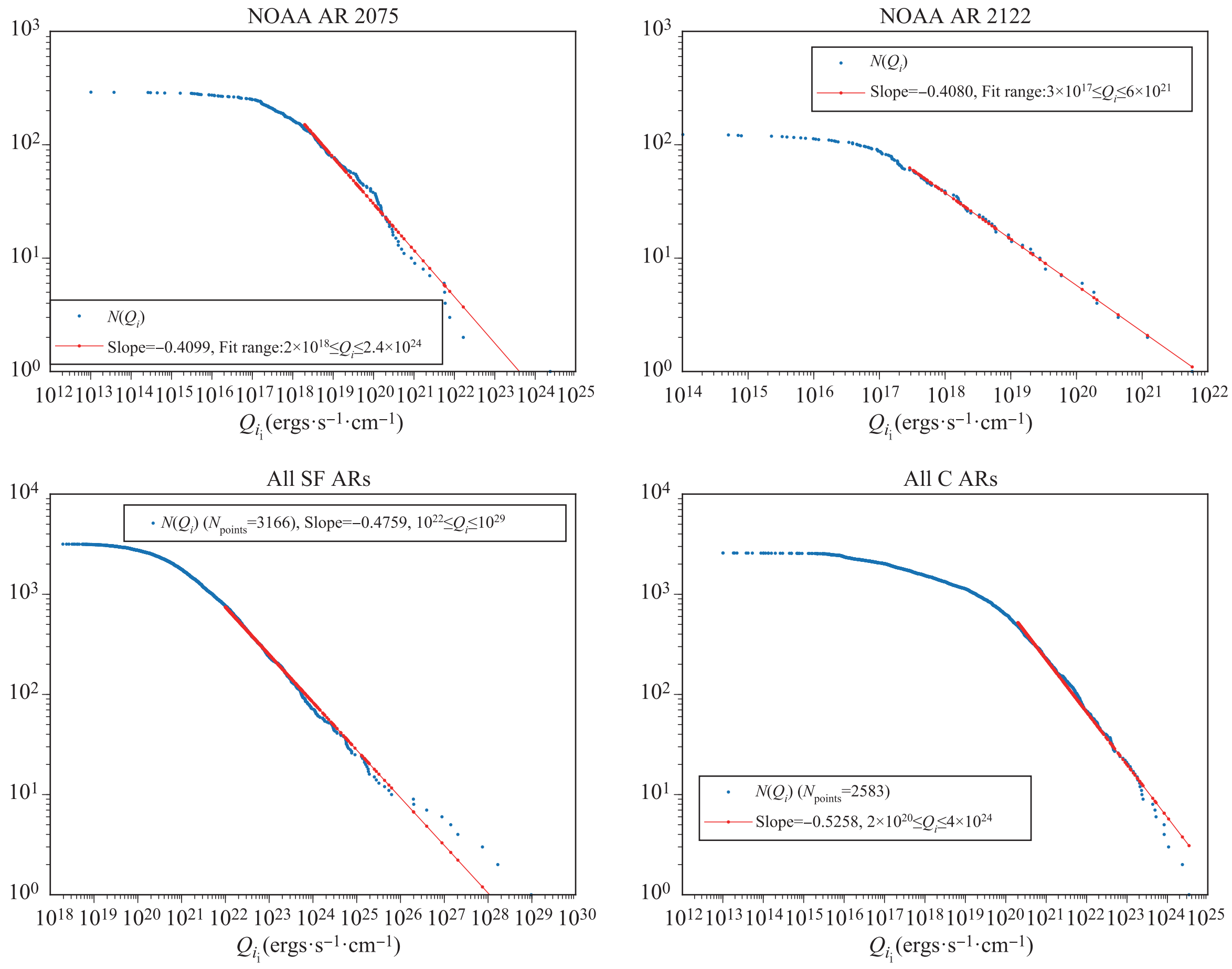

**Fig. 13** CDFs of $Q_i$ for C ARs 2075 and 2122, all SF ARs, and all C ARs. The number of orders of magnitude over which the linear fits extend are respectively 6.01, 4.3, 7, and 4.3.

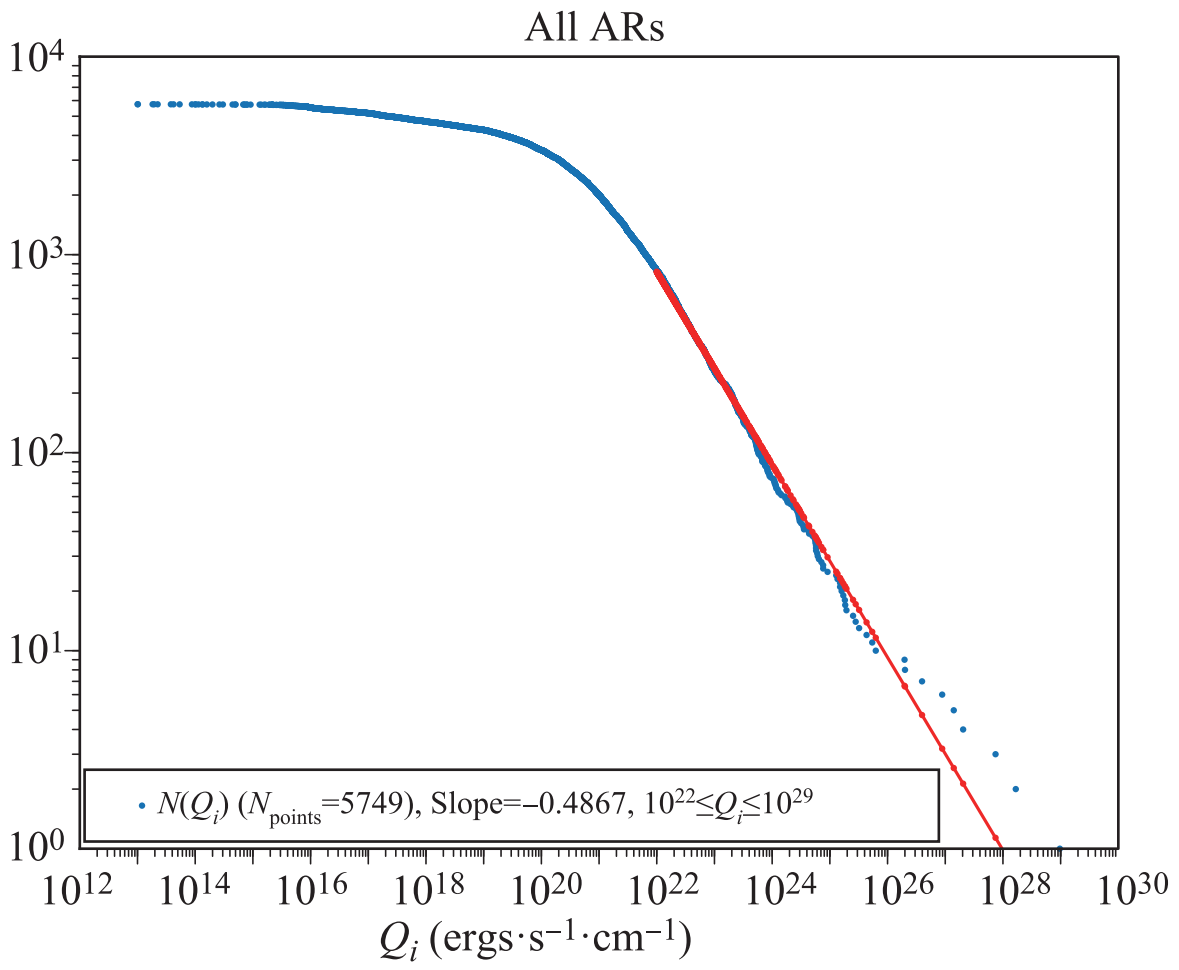

**Fig. 14** CDF of $Q_i$ for all ARs.

The local test is as follows. Several pixels in which a large current spike occurs are selected. For each such pixel the $3 \times 3$ pixel block with this pixel at its center is selected. If $t$ is the time at which the current spike occurs in the center pixel, $\boldsymbol{B}$ in each pixel in the block at times $(t-\Delta t, t, t+\Delta t)$ is determined, where $\Delta t = 12$ minutes. It is then checked if there are any reversals in the directions of $\boldsymbol{B}_h$ anywhere in the pixel block from $t - \Delta t$ to $t$, and from $t$ to $t + \Delta t$. Any such reversal indicates ambiguity error, and suggests that the current spike is due to this error. No such reversals are found for any of the spikes tested. This test is especially important because it is a direct test of whether a current spike is due to ambiguity error.

The spikes that are analyzed using the local test are found to be associated with values of $B_h$ of several hG, and values of $B_z$ orders of magnitude smaller.[8)] From Eqs. (5), (7) and (8) it follows that the horizontal current density $J_h \sim B_h|B_{x,x} + B_{y,y}|/|B_z|$, where the terms involving $B_{z,x}$ and $B_{z,y}$ are omitted since they are found not to make a major contribution to the large spikes. It is found that the larger spikes are caused by the larger values of $B_h/|B_z|$, which are due to a simultaneous occurrence of smaller values of $|B_z|$ and larger values of $B_h$. Comparably small values of $|B_z|$ occur inside and outside NLRs, and in SF and C ARs, but the values of $B_h$ associated with the larger spikes tend to be orders of magnitude larger in the NLRs of SF ARs than elsewhere.

The conclusion from the analysis in this section is that it is unlikely that ambiguity error has a significant effect on the model results. In particular, the large spikes in the heating rate $Q_i$ are not caused by ambiguity error.

As discussed in Section 5.4, the large spikes in $Q_i$ are due to spikes in $J_\perp$, and $J_\perp$ generally dominates resistive heating in the ARs, especially in the SF ARs. For the spikes analyzed in this section, the current spikes are due to spikes in $J_h$. It follows that $\boldsymbol{J}_\perp \sim \boldsymbol{J}_h$ for these spikes. This suggests that at the location of at least the large spikes, $\boldsymbol{J}$ and $\boldsymbol{B}$ are horizontal and orthogonal. At $z = 0$, Eqs. (7) and (8) imply $J_h \sim cB_h/(4\pi L_0) \sim (c/4\pi)\partial B_h/\partial z$. Together these results suggest the strong heating events occur in horizontal current sheets associated with large vertical gradients of $\boldsymbol{B}_h$ at or near the photosphere. Because the HMI spatial resolution of $\boldsymbol{B}$ is $\sim 1''$, it follows that the areas of these sheets are $\lesssim$ the characteristic area of a granule.

## 9 Data quality

The possibility that the largest current spikes are due to poor quality of the un-filtered data was investigated as follows. For each HMI image there is a data quality keyword value (henceforth "value"), with a zero value meaning good quality, and a non-zero value meaning poor quality. The values were plotted for all raw data images of $B_x$, $B_y$, and $B_z$ for the 14 AR time series analyzed. For each AR the value was checked for the images containing the two largest spikes, and sometimes additional spikes in $Q_i$. The values were all zero. The images analyzed showed a period of about 12 hours between consecutive non-zero values. This is probably due to the spurious 12 hour Doppler period.

An example of this value analysis is shown in Fig. 15. The figure shows the values for 352 time consecutive images of $B_x$ for NOAA AR 1166, including the time interval in the AR 1166 plot in Fig. 4. Of these images, 27 have nonzero values. These 27 images have image numbers 31, 32, 91, 92, 97, 98, 151, 152, 206, 207, 211, 212, 217, 218, 219, 220, 271, 272, 308, 309, 317, 318, 331, 332, 341, 342, and 343. Also shown are the image numbers of the 12 largest values of $Q_i$ in the plot for AR 1166 in Fig. 4, corresponding to $Q_i \geq 10^{24}$ ergs·s$^{-1}$·cm$^{-1}$. These 12 values correspond to image numbers 17, 24, 49, 52, 59, 106, 164, 232, 241, 260, 292, and 324. The figure shows that none of these $Q_i$ values occur at times of poor data quality. The figure also shows the approximate 12 hour periodicity of the nonzero values. The corresponding plots for $B_y$ and $B_z$ are identical to Fig. 15.

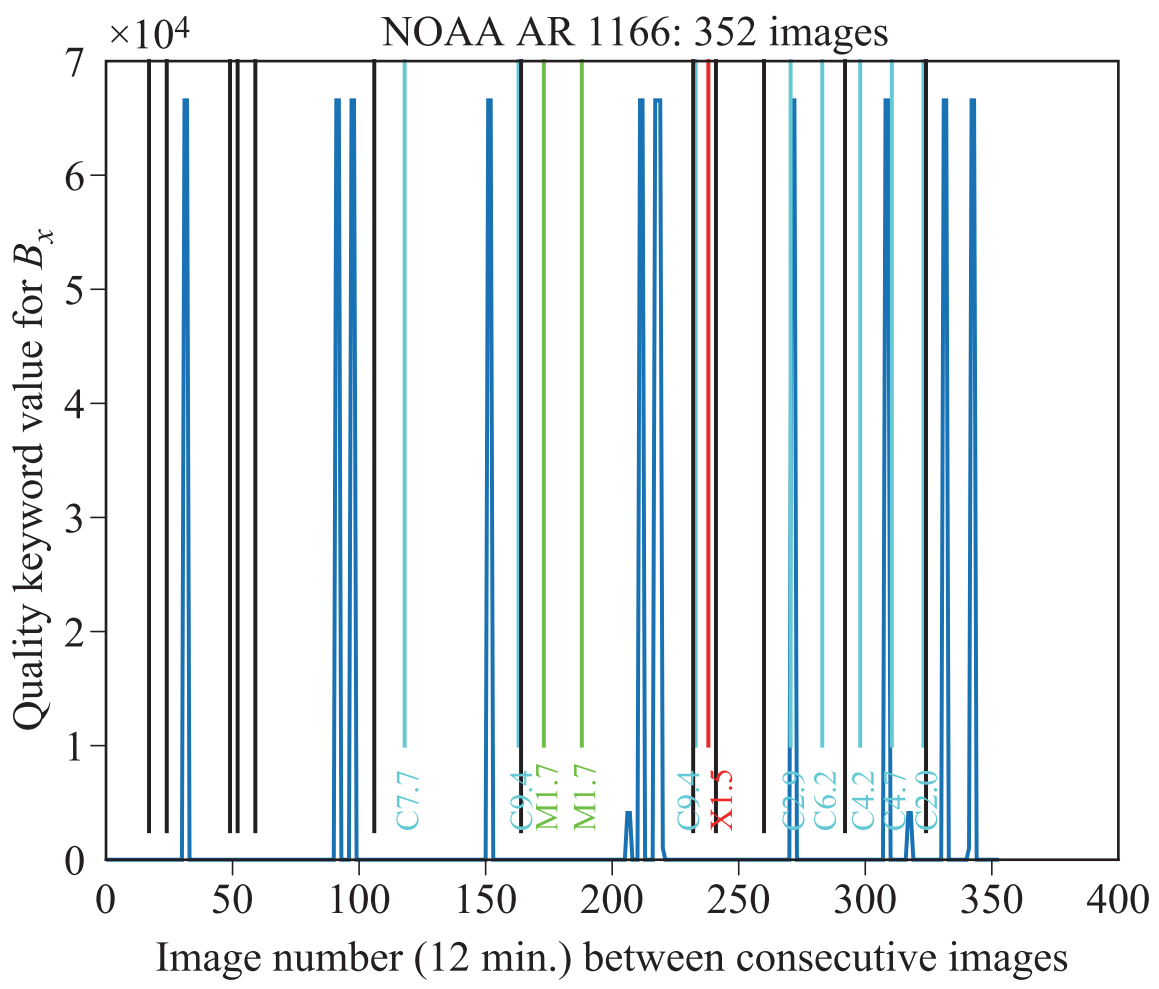

**Fig. 15** Data quality keyword values for $B_x$ in AR 1166, including the time interval shown in Fig. 4. Of these images, 27 have nonzero values, and have image numbers labeled by dark blue lines. The black lines indicate the image numbers at which the 12 largest values of $Q_i$ occur. The flare times and labels from the plot for AR 1166 in Fig. 4 are also shown.

## 10 Conclusions

The spurious Doppler induced periods in the HMI $\boldsymbol{B}$ time series can introduce significant error into photospheric quantities derived from $\boldsymbol{B}$. In particular, they can cause large errors in calculations of the photospheric $\boldsymbol{J}$, which represents free magnetic energy in the photosphere, and determines the Lorentz force $(\boldsymbol{J} \times \boldsymbol{B})/c$, and $\boldsymbol{J} \cdot \boldsymbol{E}$. They can also cause errors in the calculation of $\boldsymbol{E}$ if it is computed from $\boldsymbol{E} \sim -c^{-1}\partial \boldsymbol{A}/\partial t$, with $\boldsymbol{A}$ determined by solving $\nabla \times \boldsymbol{A} = \boldsymbol{B}$. As discussed in Section 3, as a result of filtering out these periods, our analysis is based on a high frequency representation of the photospheric magnetic field. However, in Section 5.2 it is shown there is a plausible correlation between larger values of $|\mathrm{d}\Phi_i/\mathrm{d}t|$, spikes in $Q_i$, and M and X flares. Larger values of $|\mathrm{d}\Phi_i/\mathrm{d}t|$ correspond to larger values of $|\mathrm{d}\boldsymbol{B}/\mathrm{d}t|$, which correspond to the higher frequency component of $\boldsymbol{B}$. This suggests that filtering out the lower frequency component of $\boldsymbol{B}$ retains the dynamics of $\boldsymbol{B}$ most important for flare predic-

[8)] In this paragraph the subscript 0 used in Section 2 to indicate that a quantity is evaluated at the photosphere ($z = 0$) is dropped. It is understood that all quantities are evaluated at the photosphere.

tion.

The largest photospheric heating events predicted by the model, corresponding to the largest spikes in $Q_i$ for ARs with M or X flares are plausibly correlated in time with the subsequent occurrence of M or X flares several hours to several days later.

The larger values of $Q_i$ are convection driven in that they are driven almost entirely by the conversion of bulk flow kinetic energy into thermal energy, rather than by the conversion of magnetic energy into thermal energy, at least for the time and space averaged magnetic field inferred by HMI. It is possible that reconnection initiates the energy conversion process that includes acceleration of bulk flow across magnetic field lines, which generates a convection electric field that drives current that is resistively dissipated. Higher resolution is needed to determine if this is the case.

The times of larger values of $|\mathrm{d}\Phi_i/\mathrm{d}t|$ are plausibly correlated with the subsequent occurrence of large spikes in $Q_i$, and M and X flares. Such a correlation is expected since, by Faraday's law, larger values of $|\mathrm{d}\Phi_i/\mathrm{d}t|$ imply larger values of the induction electric field, through which magnetic energy is converted into particle energy, which drives flares. By contrast, time series of the NLR integrated unsigned magnetic flux, of the magnetic energy density, and of $|\Phi_i|$ do not suggest a correlation with $Q_i$ or flares.

The CDFs of coronal flares and of $Q_i$ are essentially identical in that they are both scale invariant power law distributions, and have power law index ranges that strongly overlap and are largely independent of AR. This suggests that the processes that drive coronal flares and photospheric resistive heating are the same process operating on two largely different sets of spatial and temporal scales.

The form of the CDFs of $Q_i$ for the 14 ARs is a necessary but not sufficient condition for AR photospheres to be in an SOC state, at least outside of flaring times. For ARs that produce M or X flares, the near identity of the CDFs of $Q_i$ and coronal flares suggests that the formation of an SOC state is part of a coupled evolution of the corona and photosphere into SOC states that relax to states of lower magnetic energy through avalanches in the conversion of magnetic energy into particle energy. Results also suggest that ARs that produce weaker or no flares also evolve into an SOC state, but one that is relatively stable, consistent with the result that the spikes in $Q_i$ for these ARs are several orders of magnitude smaller than those for the ARs that produce M or X flares.

These model predicted correlations between flare times and $Q_i$ and $|\mathrm{d}\Phi_i/\mathrm{d}t|$, and the near identity of the photospheric CDF of $Q_i$ and the observed CDF of coronal flares indicate the model and model output analysis might be the basis of a new and useful algorithm for flare prediction. In order to determine the degree to which it is useful with statistical certainty, an analysis of at least hundreds of ARs involving automated feature recognition and correlation algorithms is necessary.

**Acknowledgements** This work was partially supported by a NASA Phase I SBIR award (Contract No. NNX14CG30P) to Applied Research LLC. This work made use of NASA's Astrophysics Data System (ADS). The authors are grateful to the HMI team, especially Phil Scherrer, Todd Hoeksema, Yang Liu, Monica Bobra, and Rebecca Centeno for much advice about HMI data issues. MLG thanks the Jacobs Space Exploration Group and the NASA MSFC Natural Environments Branch-EV44 for partial support of this work. The authors are also grateful to the several referees whose comments significantly improved the paper.